\documentclass[showpacs,preprintnumbers,amsmath,amssymb]{revtex4}
\usepackage{amsmath,amssymb,graphics,epsfig,subfigure}
\usepackage{color}

\begin{document}
\newcommand {\nn} {\nonumber}
\renewcommand{\baselinestretch}{1.3}

\title{Extended thermodynamics and microstructures of four-dimensional charged Gauss-Bonnet black hole in AdS space}

\author{Shao-Wen Wei \footnote{weishw@lzu.edu.cn, corresponding author},
        Yu-Xiao Liu \footnote{liuyx@lzu.edu.cn}}

\affiliation{Institute of Theoretical Physics, Lanzhou University, Lanzhou 730000, People's Republic of China\\
Joint Research Center for Physics, Lanzhou University and Qinghai Normal University, Lanzhou 730000 and Xining 810000, China}

\begin{abstract}
The discovery of new four-dimensional black hole solutions presents a new approach to understand the Gauss-Bonnet gravity in low dimensions. In this paper, we test the Gauss-Bonnet gravity by studying the phase transition and microstructures for the four-dimensional charged AdS black hole. In the extended phase space, where the cosmological constant and the Gauss-Bonnet coupling parameter are treated as thermodynamic variables, we find that the thermodynamic first law and the corresponding Smarr formula are satisfied. Both in the canonical ensemble and grand canonical ensemble, we observe the small-large black hole phase transition, which is similar to the case of the van der Walls fluid. This phase transition can also appear in the neutral black hole system. Furthermore, we construct the Ruppeiner geometry, and find that besides the attractive interaction, the repulsive interaction can also dominate among the microstructures for the small black hole with high temperature in a charged or neutral black hole system. This is quite different from the five-dimensional neutral black hole, for which only dominant attractive interaction can be found. The critical behaviors of the normalized scalar curvature are also examined. These results will shed new light into the characteristic property of four-dimensional Gauss-Bonnet gravity.
\end{abstract}



\maketitle

\section{Introduction}
\label{secIntroduction}

Since the establishment of the four black hole thermodynamic laws, phase transition has been one of the significantly active areas. In particular, black hole chemistry in the extended phase space has attracted much more attention in the past decade. One remarkable progress made in this field is that the cosmological constant is interpreted as thermodynamic pressure and its conjugate quantity as volume \cite{Kastor}. Then a full analogy between a charged anti-de Sitter (AdS) black hole system and a van der Waals fluid (VdW) was completed \cite{Kubiznak}. Subsequently, other novel phenomena of phase transition and phase structure were observed \cite{Altamirano,AltamiranoKubiznak,Altamirano3,Dolan,Frassino,
Cai,XuZhao,Kostouki,Hennigar,Hennigar2,Tjoa2,Ruihong,Ghaffarnejad}, for a recent review, see Refs. \cite{Altamiranoa,Teob}.

Black hole thermodynamics and phase transition in Gauss-Bonnet (GB) gravity, a widely concerned modified gravity, were extensively studied. Fixing the cosmological constant, it was found that the six-dimensional charged GB-AdS black hole has four extremal points at one isothermal curve, which implies that there must exist a triple point phase structure beyond the VdW one \cite{Liu}. In the extended phase space, the charged topological GB-AdS black hole was examined in Ref. \cite{Caiyang}, where the authors found that for the Ricci flat and hyperbolic GB black holes, no $P$-$V$ criticality and phase transition exists. However, the VdW-like phase transition always occurs in $d\geq 5$. Even when the charge is absence, the phase transition can also survive in five dimensions. Soon later, we clearly showed the triple point and phase structure, but it is only exists in a small parameter range for six dimensions \cite{Wei2}. Such unique property was also confirmed in Ref. \cite{Frassino}.

Exploring the black hole microstructures is always a fascinating subject. In Refs. \cite{Weiw,Weilm,Weiswliu,Weiscib}, we succeeded in developing a general approach to study black hole microstructures by using the Ruppeiner geometry \cite{Ruppeiner}---a kind of thermodynamic geometry that has important application in testing the micro-interaction of fluid systems. Comparing with the phase structure, we disclosed that different from the VdW fluid, the repulsive interaction can dominate among the black hole microstructures in any dimensions \cite{Weilm,Weiswliu}. This reveals one intriguing property of the microstructures for the charged AdS black hole. On the other hand, five-dimensional neutral GB-AdS black holes were found to have an analytical coexistence curve of small and large black holes \cite{Mol}, which provides a good opportunity to exactly test the black hole microstructures in modified gravity. Nevertheless, it was found that the dominant repulsive interaction is absent. However, the scalar curvature keeps constant when the black hole system undergoes a phase transition \cite{Weiplb}. That means the interaction of the five dimensional neutral GB-AdS black hole maintains the same even when its microstructures have an obvious change. It uncovers a novel property for the black hole in the modified gravity.

As we know, in GB gravity, only the static and spherically symmetric black hole solutions are found in higher dimensions. While in four dimensions, the GB term is a total derivative and it has no influence on the field equation, which results in no 4D GB black hole solution exists. In Refs. \cite{Tomozawa,Cognola}, the authors considered the GB gravity in four dimensions by rescaling the GB coupling parameter $\alpha\rightarrow\frac{\alpha}{d-4}$. Recently, Glavan and Lin \cite{Glavan} took the similar technique and proposed a general covariant GB modified gravity in four dimensions. Such theory can bypass the Lovelock's theorem and avoid Ostrogradsky instability. After taking the limit $d\rightarrow$4, the GB term has nontrivial contribution to the field equation. A non-trivial and novel four-dimensional static and spherically symmetric black hole solution was obtained \cite{Glavan}. The quasinormal modes of the solution were examined in \cite{Zinhailo}, where the result shows that the damping rate is more sensitive than the real part by varying the GB coupling parameter. The shadow size of the GB black hole was also investigated in Refs. \cite{Zinhailo,Guoli}. The negative GB coupling enlarges the shadow, while the positive one shrinks it \cite{Guoli}. The four-dimensional charged GB-AdS black hole was also found in Ref. \cite{Fernandes}. Subsequently, other related topics have also been studied \cite{Casalino,Konoplyad,WeiGB,Kumarsg,Hegde,Doneva,Zhangyp,Lu,Singh,Konoplyas,Ghoshku,Zhidenkoko,Kobayashi,KumarKumar,Zhangli,Mansoori}.

Thermodynamics is an important aspect of a black hole solution. Moreover, as shown above,  the higher-dimensional GB black holes have a novel phase transition and microstructures. In this paper, we would like to examine the corresponding properties in four dimensions and see whether these results still hold or not. We believe the study can help us to further uncover the underlying interesting properties of the four-dimensional GB gravity.

The present paper is organized as follows. In Sec. \ref{dddd}, we study the thermodynamics of the four-dimensional charged GB-AdS black hole. The thermodynamic first law and the Smarr formula are checked. In Sec. \ref{sss}, the phase transition is investigated in the canonical ensemble. We observe a small-large black hole phase transition. The critical point and critical exponent are calculated. We also address the issue that different choice of the entropy will affect the coexistence curve. And the inconsistency between these curves obtained from the equal area law and Gibbs free energy will emerge. In the grand canonical ensemble, see Sec. \ref{se}, we also observe the small-large black hole phase transition. Interestingly, in the reduced parameter space, all the thermodynamic quantities and coexistence curve are independent of the GB coupling parameter. In Sec. \ref{bhm}, we construct the thermodynamic geometry. The normalized scalar curvature is calculated. The possible property of the microstructure is uncovered. Especially, the critical phenomena of the scalar curvature is numerically examined. Finally, we summarize and discuss our results in Sec. \ref{Conclusion}.

\section{Black hole thermodynamics}
\label{dddd}

It is well known that, in GB gravity, there are the static and spherically symmetric black hole solutions with $d\geq5$ \cite{Boulware,Wiltshire,Cairg,Nojiria,Cvetica}. Since the GB term is a total derivative and has no contribution to the field equation in four-dimensional spacetime, there is no four-dimensional GB black hole. However, as discovered by Glavan and Lin \cite{Glavan}, when rescaling the GB coupling parameter $\alpha\rightarrow\alpha/(d-4)$, then taking the limit $d\rightarrow4$, they obtained a four-dimensional non-trivial black hole solution. Subsequently, this black hole solution was generalized to charged case in an AdS space \cite{Fernandes}.

The action of the Gauss-Bonnet gravity in a $d$-dimensional spacetime with a negative cosmological constant is
\begin{eqnarray}
 &&S=\frac{1}{16\pi}\int d^dx\left(R+\frac{(d-1)(d-2)}{l^2}+\frac{\alpha}{d-4}L_{GB}-F_{\mu\nu}F^{\mu\nu}\right),
 \end{eqnarray}
where
 \begin{eqnarray}
 &&L_{GB}=R_{\mu\nu\rho\sigma}R^{\mu\nu\rho\sigma}-4R_{\mu\nu}R^{\mu\nu}+R^2,
\end{eqnarray}
the AdS radius $l$ is related to the cosmological constant as $\Lambda=-(d-1)(d-2)/(2l^2)$, and $F_{\mu\nu}$ is the Maxwell tensor. The black hole solution was obtained by solving the field equation and adopting the limit $d\rightarrow4$ in Ref.~\cite{Fernandes}:
\begin{eqnarray}
 ds^2&=&-f(r)dt^2+\frac{1}{f(r)}dr^2+r^2(d\theta^2+\sin^2\theta d\phi^2),\\
 f(r)&=&1+\frac{r^2}{2\alpha}\left(1-\sqrt{1+4\alpha\left(\frac{2M}{r^3}-\frac{Q^2}{r^4}-\frac{1}{l^2}\right)}\right),
\end{eqnarray}
with the non-vanishing electrostatic vector potential $A_t=Q/r$. The parameters $M$ and $Q$ are the black hole mass and charge, respectively. Taking the limit $\alpha\rightarrow0$, the Reissner-Nordstr\"om-AdS will be recovered. The static and spherically symmetric black hole solution found in Ref.~\cite{Glavan} can be obtained by setting $Q=0$ and $l\rightarrow\infty$. It is worth to point out that this GB black hole solution has the same form with the one obtained in a conformal anomaly gravity \cite{caicao,cai3}, while the parameter has different meanings. The radius $r_{h}$ of the black hole horizon is the largest root of $f(r)=0$. Making use of it, we can express the black hole mass as
\begin{eqnarray}
 M=\frac{3 \alpha +8 \pi  P r_h^4+3r_h^2+3 Q^2}{6 r_h},
\end{eqnarray}
with the interpretation that the cosmological constant is the pressure \cite{Kastor}
\begin{eqnarray}
 P=\frac{3}{8 \pi l^2}.
\end{eqnarray}
The temperature can be calculated as
\begin{eqnarray}
 T=\frac{8 \pi  P r_h^4+r_h^2-Q^2-\alpha}{8 \pi  \alpha
   r_h+4 \pi  r_h^3}.\label{tempe}
\end{eqnarray}
As shown in Refs. \cite{Caiyang,Wei2}, the GB coupling parameter $\alpha$ is also treated as a new thermodynamic variable. Then the black hole first law reads
\begin{eqnarray}
 dM=TdS+\Phi dQ+VdP+\mathcal{A}d\alpha.\label{enthalpy}
\end{eqnarray}
Here $\Phi$ denotes the electrostatic potential on the horizon, i.e., $\Phi=A_t(r_h)=Q/r_h$. The parameters $V$ and $\mathcal{A}$ are the conjugate quantities of the pressure $P$ and GB coupling parameter $\alpha$, respectively. After the entropy is obtained, these quantities can be easily calculated by using the first law. Following \cite{Cairgsoh}, for fixed $Q$, $P$, and $\alpha$, the black hole entropy is obtained \cite{Fernandes}
\begin{eqnarray}
 S=\int\frac{dM}{T}&=&\int\frac{1}{T}\left(\frac{\partial M}{\partial r_h}\right)dr_h\nonumber\\
 &=&\frac{A}{4}+2 \pi  \alpha  \ln \left(\frac{A}{A_0}\right),
\end{eqnarray}
where $A=4\pi r_h^2$ is the area of the black hole horizon and $A_0$ is an integral constant with dimension of [length]$^2$. Note that this entropy is the same as that obtained from the Iyer-Wald formula \cite{Lu}. It is also clear that the Bekenstein-Hawking entropy-area law is modified by the GB coupling parameter $\alpha$. As we know, for the black hole, the entropy is generally independent of the black hole charge and the cosmological constant, so we set $A_0=4\pi |\alpha|$ for simplicity. After the choice, the entropy has the following form
\begin{eqnarray}
 S=\pi  r_h^2+4 \pi \alpha\ln\left(\frac{r_h}{\sqrt{|\alpha|}}\right).\label{entro}
\end{eqnarray}
Note that the modified term of our entropy is different from that of \cite{Hegde}, where the integral constant is neglected. It is worth pointing out that different entropy will produce different phase transition point from the Gibbs free energy, as well as the phase diagram and coexistence curves. In the following, we only focus on positive $\alpha$. The thermodynamic volume $V$ and the conjugate quantity $\mathcal{A}$ to $\alpha$ are
\begin{eqnarray}
 V&=&\left(\frac{\partial M}{\partial P}\right)_{S, Q, \alpha}=\frac{4}{3}\pi r_h^3,\\
 \mathcal{A}&=&\left(\frac{\partial M}{\partial \alpha}\right)_{S, Q, P}=\frac{\alpha +2 \ln
   \left(\frac{r_h}{\sqrt{\alpha
   }}\right) \left(\alpha -8 \pi  P
   r_h^4-r_h^2+Q^2\right)+8 \pi  P
   r_h^4+2 r_h^2-Q^2}{2 \left(2 \alpha
   r_h+r_h^3\right)}.
\end{eqnarray}
On the other hand, from the expression (\ref{enthalpy}), one can clearly see that the black hole mass now plays the role of the enthalpy of the system, i.e., $H\equiv M$. Further, with these thermodynamic quantities, we confirm the following Smarr formula
\begin{eqnarray}
 H=2TS+\Phi Q-2PV+2 \alpha\mathcal{A}.
\end{eqnarray}
Moreover, the first law (\ref{enthalpy}) can also be checked.

\section{Phase transition in canonical ensemble}
\label{sss}

In this section, we would like to study the phase transition in the canonical ensemble, where the black hole charge $Q$ is fixed.

\begin{figure}
\center{\includegraphics[width=8cm]{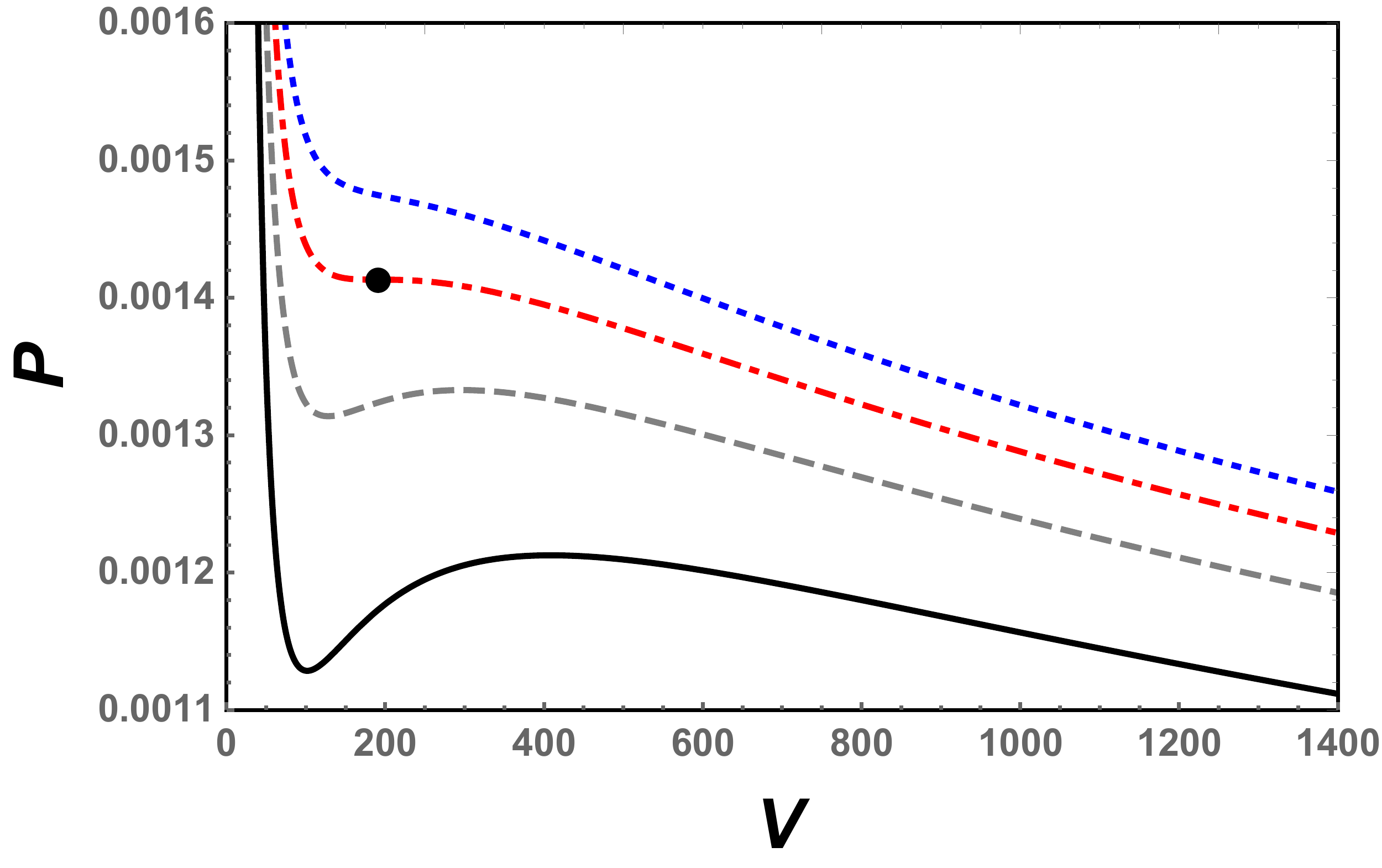}}
\caption{Isothermal curves for the charged GB-AdS black hole with $Q$=1 and $\alpha$=0.5. The temperature is taken as $T$=0.026, 0.027, 0.02759 ($T_c$), and 0.028 from bottom to top. The black dot at $V$=189.76 denotes a critical point.}\label{ppA0rsa}
\end{figure}

Solving (\ref{tempe}) for the pressure, we can obtain the state equation for the four-dimensional charged GB-AdS black hole
\begin{eqnarray}
 P=\frac{T}{2 r_h}+\frac{\alpha T}{r_h^3}-\frac{1}{8\pi  r_h^2}+\frac{Q^2+\alpha}{8 \pi  r_h^4}. \label{P-T}
\end{eqnarray}
Comparing with the state equation of VdW fluid $P=T/v+\cdots$, one can define a specific volume $v=2r_h$ for the charged AdS black hole \cite{Kubiznak}. However, for our case, the appearance of the second term in Eq.~(\ref{P-T}) makes the specific volume more difficult to understand. So we will abandon the concept. On the other hand, employing the thermodynamic volume $V$, the state equation will be
\begin{eqnarray}
 P=\frac{4 \pi \alpha T}{3 V}-\frac{1}{2\times 6^{2/3} \sqrt[3]{\pi} V^{2/3}}+\frac{\sqrt[3]{\frac{\pi }{6}} \left(\alpha +Q^2+3 T V\right)}{3V^{4/3}}\label{ndp}.
\end{eqnarray}
We plot the isothermal curves in Fig. \ref{ppA0rsa} for $Q$=1 and $\alpha$=0.5. From bottom to top, the temperature is taken as $T$=0.026, 0.027, 0.02759 ($T_c$), and 0.028, respectively. For low temperature, there are two extremal points dividing the system to three branches---small, intermediate and large black holes characterized by different values of the thermodynamic volume. The small and large black hole branches have positive heat capacity, which implies that they are thermodynamically stable. However the intermediate one with negative heat capacity is unstable. This branch will be removed according to the Maxwell equal area law. And a small-large black hole phase transition occurs reminiscent of the liquid-gas phase transition of the VdW fluid. Increasing the temperature, the two extremal points get closer, and coincide with each other at the critical point, beyond which the small and large black hole branches will not be clearly distinguished.

\begin{figure}
\center{\subfigure[]{\label{PTcQa}
\includegraphics[width=7cm]{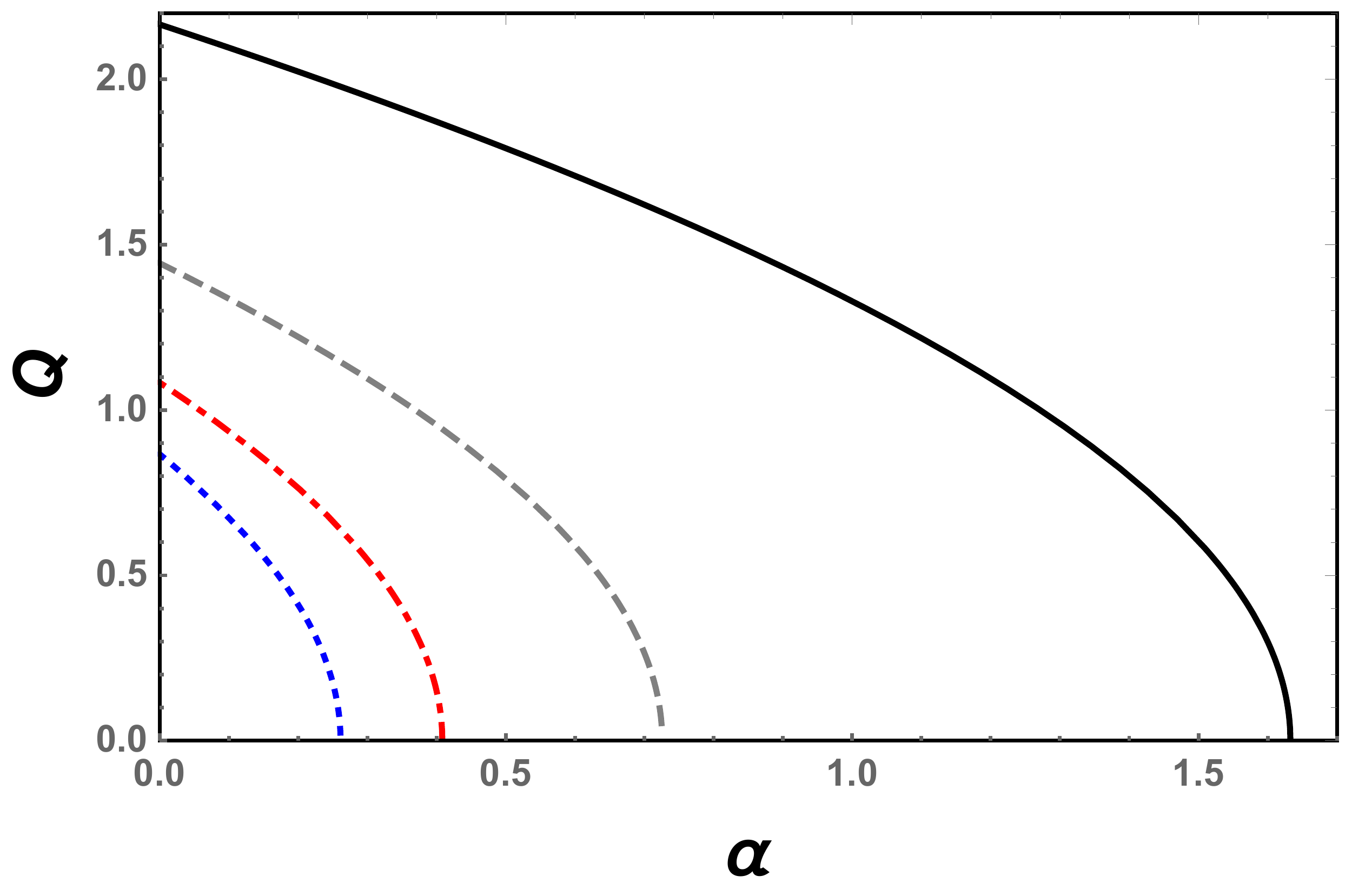}}
\subfigure[]{\label{PPcQa}
\includegraphics[width=7cm]{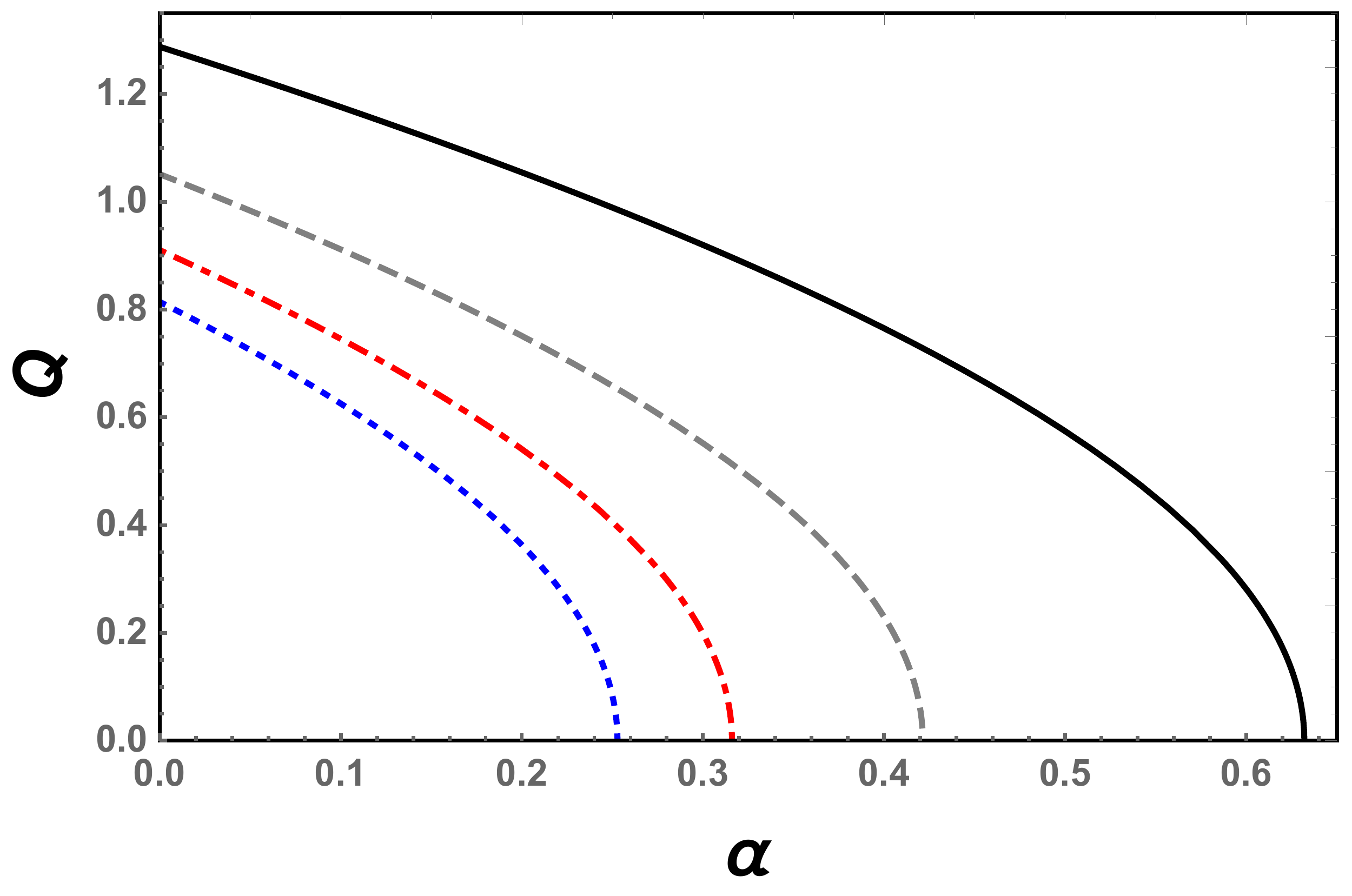}}}
\caption{Contours of the critical temperature and pressure in the $Q$-$\alpha$ diagram. (a) Contours of the critical temperature $T_c$=0.02, 0.03, 0.04, and 0.05 from right to left. (b) Contours of the critical pressure $P_c$=0.002, 0.003, 0.004, and 0.005 from right to left. }\label{ppPcQa}
\end{figure}

The critical point denoting a second-order phase transition is determined by the following equations
\begin{eqnarray}
 \left(\frac{\partial P}{\partial V}\right)_{T, Q, \alpha}=0,\quad
 \left(\frac{\partial^2 P}{\partial V^2}\right)_{T, Q, \alpha}=0.\label{crcond}
\end{eqnarray}
Solving them, one gets \cite{Hegde}
\begin{eqnarray}
 T_c&=&\frac{\kappa-3 Q^2}{24 \pi  \alpha
   \sqrt{6 \alpha +3 Q^2+\kappa}},\label{c1}\\
  P_c&=&\frac{9 \alpha +6 Q^2+\kappa}{24 \pi
   \left(6 \alpha +3 Q^2+\kappa\right)^2},\\
  V_c&=&\frac{4\pi}{3}\left(6 \alpha +3 Q^2+\kappa\right)^{\frac{3}{2}},\label{c3}
\end{eqnarray}
with $\kappa=\sqrt{48 \alpha ^2+9 Q^4+48 \alpha  Q^2}$. Obviously, the critical point is closely dependent on the charge $Q$ and GB coupling parameter $\alpha$. In order to see their effects on the critical point, we show the contours of $T_c$ and $P_c$ in Fig. \ref{ppPcQa} in the $Q$-$\alpha$ diagram. From the figures, we find that with the increase of the critical temperature or pressure, the contours are shifted to small $Q$ and $\alpha$. Moreover, for each contour, the charge $Q$ decreases with $\alpha$, which indicates that they compete with each other in this charged GB-AdS black hole system.

When setting $\alpha$=0, the values of the critical point (\ref{c1})-(\ref{c3}) reduce to that of the charged AdS black hole. While taking $Q$=0, we have
\begin{eqnarray}
 T_c&=&\frac{\sqrt{2 \sqrt{3}-3}}{6 \pi  \sqrt{2\alpha }},\label{ads1}\\
  P_c&=&\frac{15-8 \sqrt{3}}{288 \pi  \alpha },\\
  V_c&=&\frac{4\pi}{3}\left(4 \sqrt{3} \alpha +6\alpha \right)^{3/2}.\label{ads2}
\end{eqnarray}
This result indicates that the four-dimensional neutral GB-AdS black hole also has a critical point, and thus the VdW type phase transition can be found in the neutral case. Meanwhile, for the VdW fluid, by using he critical thermodynamic quantities, one can construct a universal dimensionless quantity $\rho=P_cv_c/T_c$ with $v_c$ the critical specific volume. It is found that $\rho=3/8$, which is independent of the two characteristic parameters $a$ and $b$ of the VdW fluid. So this reveals a universal property for these fluid systems modeled by the VdW model. Similarly, in Ref. \cite{Kubiznak}, the authors calculated it for the four-dimensional charged AdS black hole, and found a charge independent value $\rho=3/8$, which indicates a universal property of the black hole system. Here we show it for the charged GB-AdS black hole
\begin{eqnarray}
 \rho=\frac{2P_cr_{hc}}{T_c}=\frac{24 \alpha +21 Q^2-\kappa}{48 \left(\alpha +Q^2\right)}.\label{univq}
\end{eqnarray}
For the charged black hole with $\alpha$=0, we have $\rho=3/8$ as expected. While for the neutral GB-AdS black hole, we have a universal quantity $\rho=\left(6-\sqrt{3}\right)/12$, which is smaller than that of the charged black hole case. However, for the case of nonvanishing $\alpha$ and $Q$, we find that $\rho$ depends both on $\alpha$ and $Q$. So it seems that $\rho$ could not reflect the universal property of the black hole for this case.

Next, we turn to the first-order phase transition. It can be determined by the equal area law along the isothermal curve or swallow tail behavior of the Gibbs free energy. Here we focus on the latter one first. The corresponding Gibbs free energy in the canonical ensemble reads
\begin{eqnarray}
 G=H-TS=\frac{1}{6 r_h}\left(3 \alpha +r_h^2 \left(8 \pi  P r_h^2-6 \pi  T r_h+3\right)-24 \pi \alpha T r_h \log \left(\frac{r_h}{\sqrt{\alpha}}\right)+3 Q^2\right).\label{gibbs}
\end{eqnarray}
Here we would like to present a comment on the different choices of the entropy on determining the first-order black hole phase transition. For the same pressure, temperature, and thermodynamic volume, we can obtain the phase transition point following the equal area law. This approach is independent of the choice of entropy, and is appropriately shown in Ref. \cite{Weiprd2015}. On the other hand, it is clear that the Gibbs free energy (\ref{gibbs}) depends on the entropy. So different entropy produces different Gibbs free energy, and its swallow tail behavior, as well as the phase transition, will be influenced. As a result, the phase transition point obtained from the equal area law and Gibbs free energy may be not consistent with each other. Following the discussion of Ref. \cite{Weiprd2015}, these two approaches give the same result only when the first law (\ref{enthalpy}) holds.

\begin{figure}
\center{\includegraphics[width=8cm]{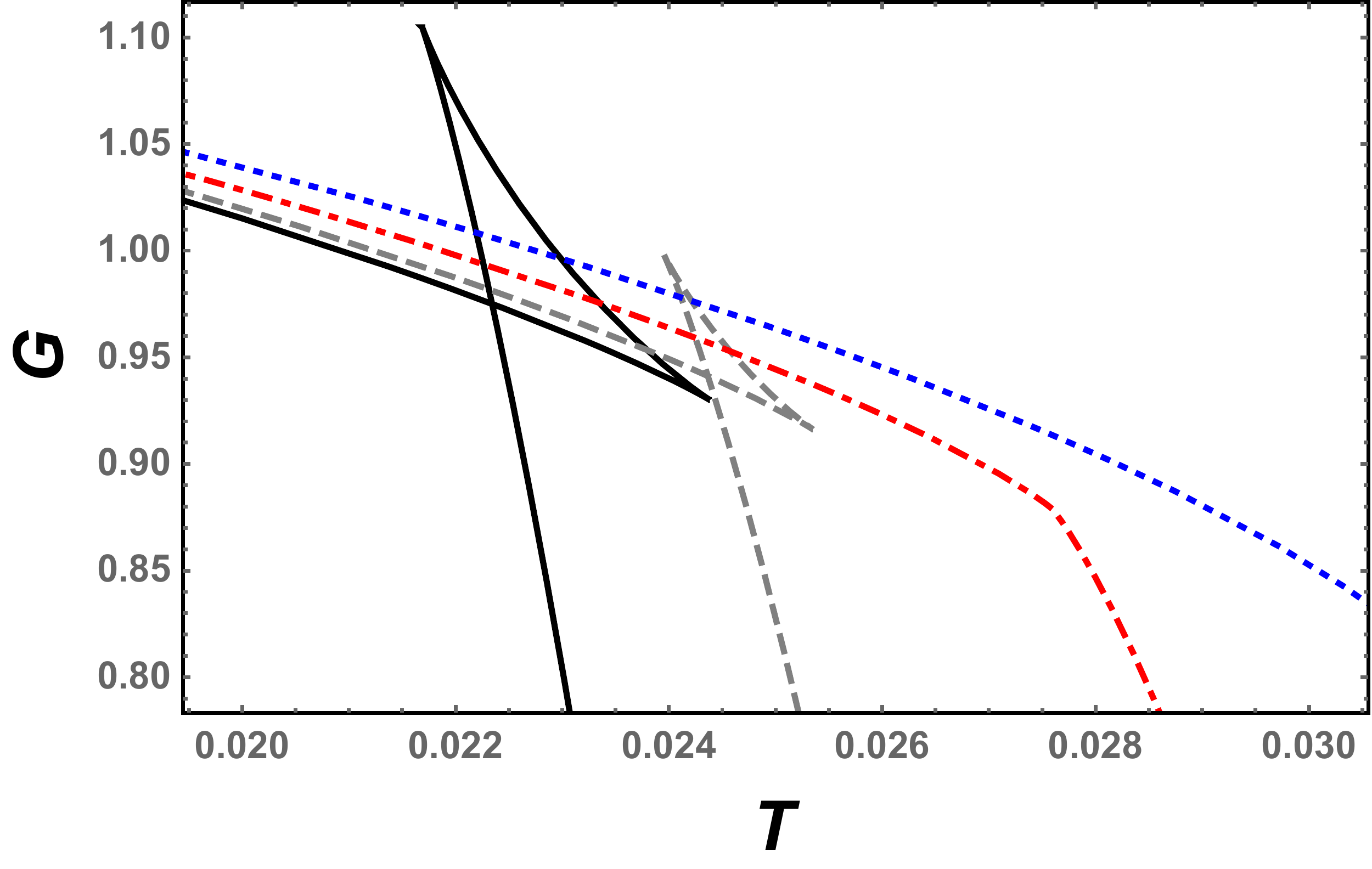}}
\caption{Gibbs free energy $G$ as a function of the temperature $T$ with $Q$=1 and $\alpha$=-0.5. The pressure $P$=0.0008, 0.0010, 0.0014 ($P_c$), 0.0020 from left to right, respectively.}\label{ppGibbs}
\end{figure}

In order to show the behavior of the Gibbs free energy, we describe it in Fig. \ref{ppGibbs} with $Q$=1 and $\alpha$=0.5 for different values of the pressure. For low pressure, there appears the swallow tail behavior. At the critical pressure, this characterized behavior disappears. When beyond the critical point, Gibbs free energy is a decreasing function of the temperature, and thus no phase transition exists. Reading out the self-cross point, one can obtain the phase transition temperature and pressure.

\begin{figure}
\center{\subfigure[]{\label{PPTa}
\includegraphics[width=7cm]{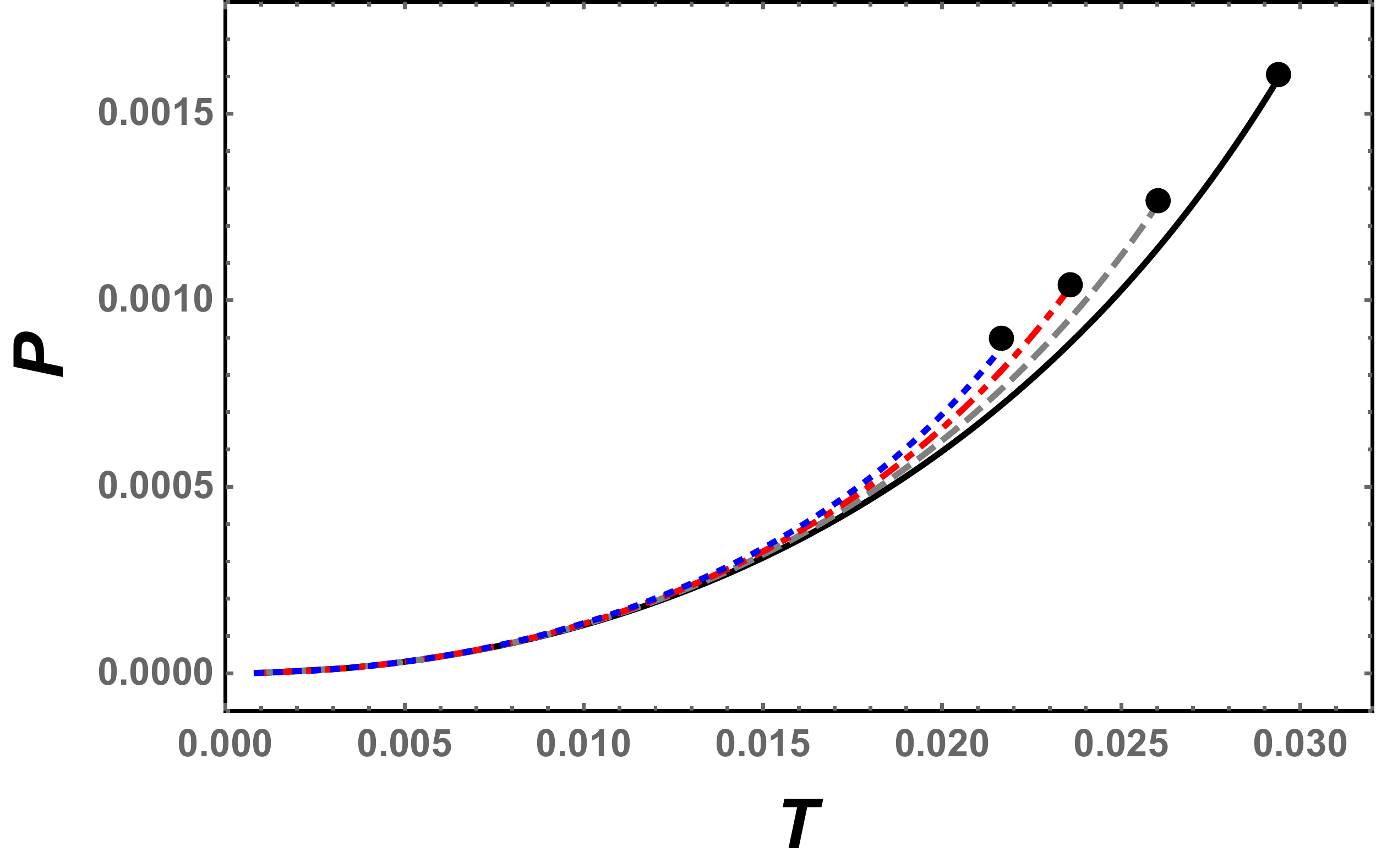}}
\subfigure[]{\label{PPT04}
\includegraphics[width=7cm]{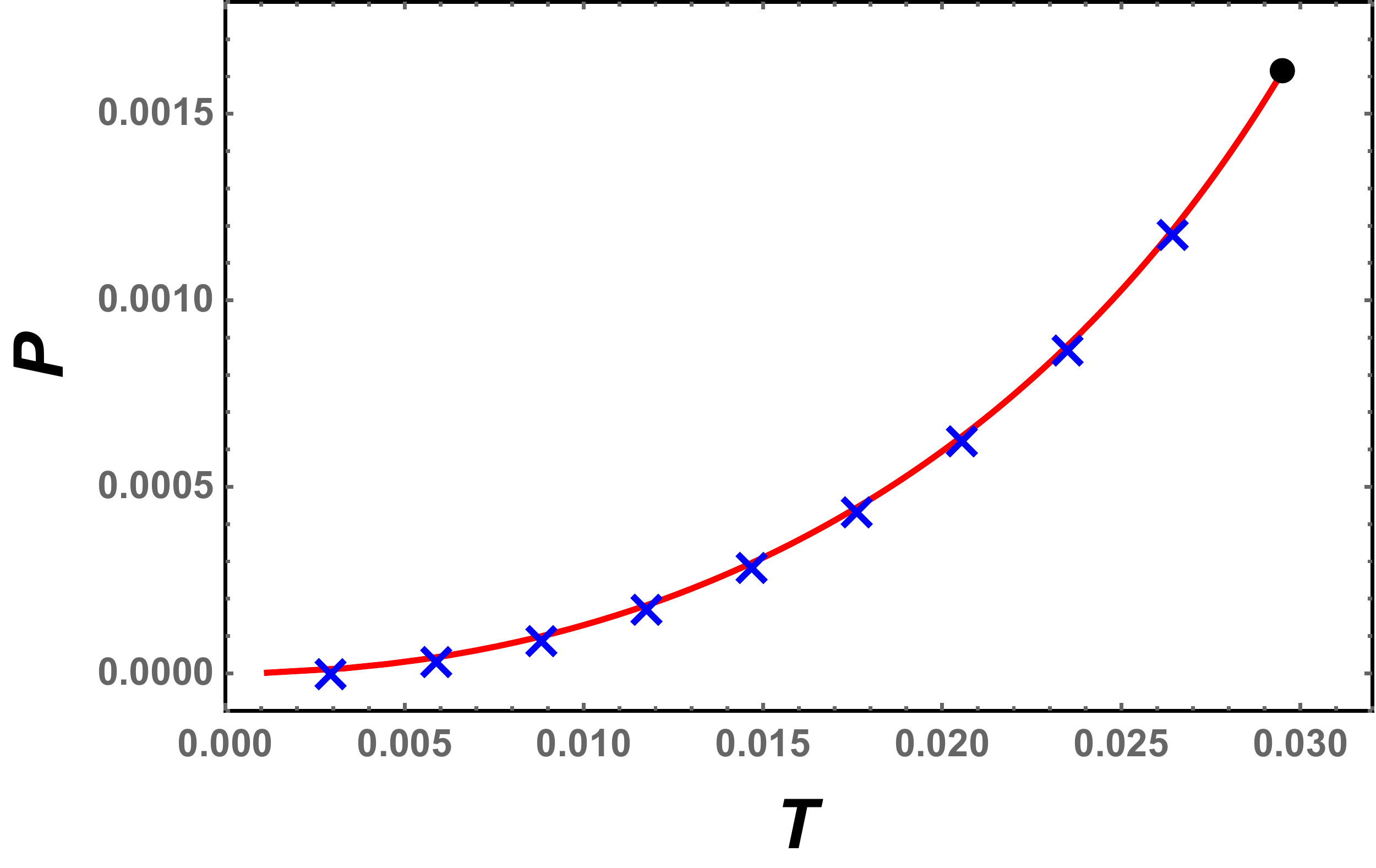}}}
\caption{$P$-$T$ phase diagram with charge $Q$=1. Black dots denote the critical points. The coexistence curves are obtained from the Gibbs free energy. (a) $\alpha$=0.4, 0.6, 0.8, and 1.0 from right to left. (b) $\alpha$=0.4. The ``$\times$" represents the result calculated from the equal area law.}\label{ppPT04}
\end{figure}

Here we show the coexistence curve in Fig. \ref{PPTa} with the charge $Q$=1. Varying $\alpha$ from 0.4 to 1.0, the coexistence curves are described from right to the left. For low temperature or pressure, we can see that these curves are well coincides with each other. However, when the critical points (marked with black dots) are approached, they deviate from each other. Obviously, with the increase of $\alpha$, the critical point is shifted to low temperature and pressure.

To check the consistency of the coexistence curves obtained from the Gibbs free energy and equal area law, we take $\alpha$=0.4 for example, see Fig. \ref{PPT04}. The red solid curve is obtained from the swallow tail behavior, while the symbol ``$\times$" denotes that calculated from the equal area law. From the figure, we find that they are well consistent with each other. If one neglects the integral constant $A_0$ in (\ref{entro}), the coexistence curves obtained from these two methods will deviate from each other.
So, the consistency shown in Fig. \ref{PPT04} indicates that our choice of the integral constant $A_0$ is appropriate. On the other hand, there might have other choice, but one needs to ensure that both the first law and the Smarr formula are satisfied.

Next, we will compute the critical exponent, which is a universal property of the phase transition. Usually, near the critical point, a VdW-like phase transition is characterized by the four critical exponents $\alpha'$, $\beta$, $\gamma$, and $\delta$, which are defined as:
\begin{eqnarray}
 C_V&=&T\left(\frac{\partial S}{\partial T}\right)_{V}\propto |t|^{-\alpha'},\\
 \eta&=&\frac{V_l-V_s}{V_c}\propto |t|^{\beta},\\
 \kappa_T&=&-\frac{1}{V}\left(\frac{\partial V}{\partial P}\right)_{T}\propto |t|^{-\gamma},\\
 p&\propto&\omega^{\delta},
\end{eqnarray}
with
\begin{eqnarray}
 p=\frac{P}{P_c}, \quad t=\frac{T-T_c}{T_c},\quad
 \omega=\frac{V-V_c}{V_c}.
\end{eqnarray}
We start with the exponent $\alpha'$, which governs the behavior of the specific heat at constant volume. It is easy to find that for fixed $\alpha$ and $V$, $(\partial_ST)_{V,\alpha}$=0. Thus we have $C_{V, \alpha}$=0, and arrive the conclusion that the exponent $\alpha'$=0. Other exponents can be calculated by expanding the state equation near the critical point, which gives
\begin{eqnarray}
 p=1+a_{10}t+a_{11}t\omega+a_{03}\omega^3+\mathcal{O}(t\omega^2, \omega^4).\label{csel}
\end{eqnarray}
The coefficients for this four-dimensional charged GB-AdS black hole system are given by
\begin{eqnarray}
 a_{10}=\frac{4 \left(2 \alpha +\kappa +3
   Q^2\right)}{11 \alpha +9 Q^2},\quad
  a_{11}=\frac{24 \alpha -10 \kappa +6 Q^2}{33 \alpha
   +27 Q^2},\quad
   a_{03}=\frac{-20 \alpha +\kappa -15 Q^2}{27 \left(11
   \alpha +9 Q^2\right)}.
\end{eqnarray}
When the black hole system undergoes a phase transition from a small black hole to a large one, the temperature and pressure keep constant, while the thermodynamic volume changes from $\omega_s$ to $\omega_l$. Before and after the phase transition, the state equation always holds
\begin{eqnarray}
 p&=&1+a_{10}t+a_{11}t\omega_s+a_{03}\omega_s^3\nonumber\\
  &=&1+a_{10}t+a_{11}t\omega_l+a_{03}\omega_l^3.
\end{eqnarray}
A simple calculation gives
\begin{eqnarray}
 a_{11}t(\omega_l-\omega_s)+a_{03}(\omega^3_l-\omega^3_s)=0.\label{eq1}
\end{eqnarray}
Moreover, during the phase transition, the Maxwell’s area law also holds
\begin{eqnarray}
 \int_{\omega_s}^{\omega^l}\omega\frac{dp}{d\omega}d\omega=0,
\end{eqnarray}
which reduces to
\begin{eqnarray}
 a_{11}t(\omega^2_l-\omega^2_s)+\frac{3}{2}a_{03}(\omega^4_l-\omega^4_s)=0.\label{eq2}
\end{eqnarray}
Solving (\ref{eq1}) and (\ref{eq2}), we obtain a non-trivial solution
\begin{eqnarray}
 \omega_l=-\omega_s=\frac{3 \sqrt{24 \alpha -10 \kappa +6
   Q^2}}{\sqrt{-20 \alpha +\kappa -15 Q^2}}\sqrt{-t}.
\end{eqnarray}
Then the order parameter $\eta$ reads
\begin{eqnarray}
 \eta=2\omega_l\propto\sqrt{-t},
\end{eqnarray}
so the critical exponent $\beta=\frac{1}{2}$. The isothermal compressibility $\kappa_T$ has
\begin{eqnarray}
 \kappa_T\propto-\left(\frac{\partial p}{\partial \omega}\right)^{-1}\sim-\frac{1}{a_{11}t},
\end{eqnarray}
which implies the exponent $\gamma$=1. The shape of the critical isotherm at $t$ = 0 is
\begin{eqnarray}
 p-1=a_{03}\omega^{3}.
\end{eqnarray}
This gives $\delta$=3. In summary, we obtain these four critical exponents
\begin{eqnarray}
 \alpha'=0,\quad \beta=\frac{1}{2},\quad\gamma=1, \quad \delta=3.
\end{eqnarray}
Actually, these exponents are not independent from each other, and they satisfy the following scaling laws
\begin{eqnarray}
 &\alpha'+2\beta+\gamma=2,\quad \alpha'+\beta(1+\delta)=2,\\
 &\gamma(1+\delta)=(2-\alpha')(\delta-1),\quad\gamma=\beta(\delta-1).
\end{eqnarray}
Now it is rather clear that the critical exponents and the scaling laws for the four-dimensional charged GB-AdS black hole are exactly the same with the mean field theory. And this small-large black hole phase transition is the VdW-like type.

\section{Phase transition in grand canonical ensemble}
\label{se}

In this section, we focus on the black hole thermodynamics and phase transition in the grand canonical ensemble, where the electric potential rather the charge is fixed. Moreover, we only consider the phase transition between two black holes, so we will not consider the pure thermal radiation in AdS space. For simplicity, we only consider the case of positive electric potential.

\begin{figure}
\center{\subfigure[]{\label{ppGPV}
\includegraphics[width=7cm]{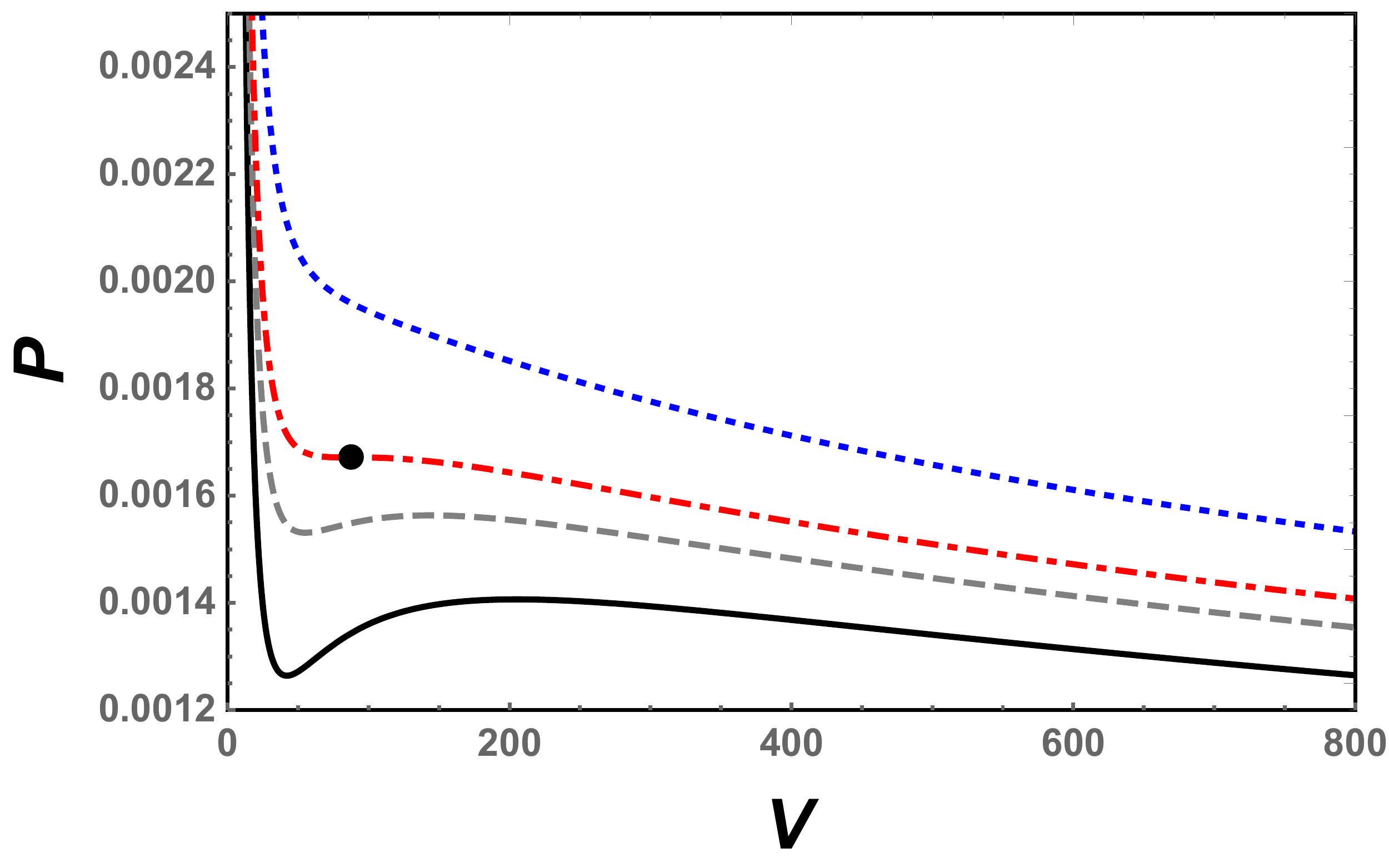}}
\subfigure[]{\label{PPGranFT}
\includegraphics[width=7cm]{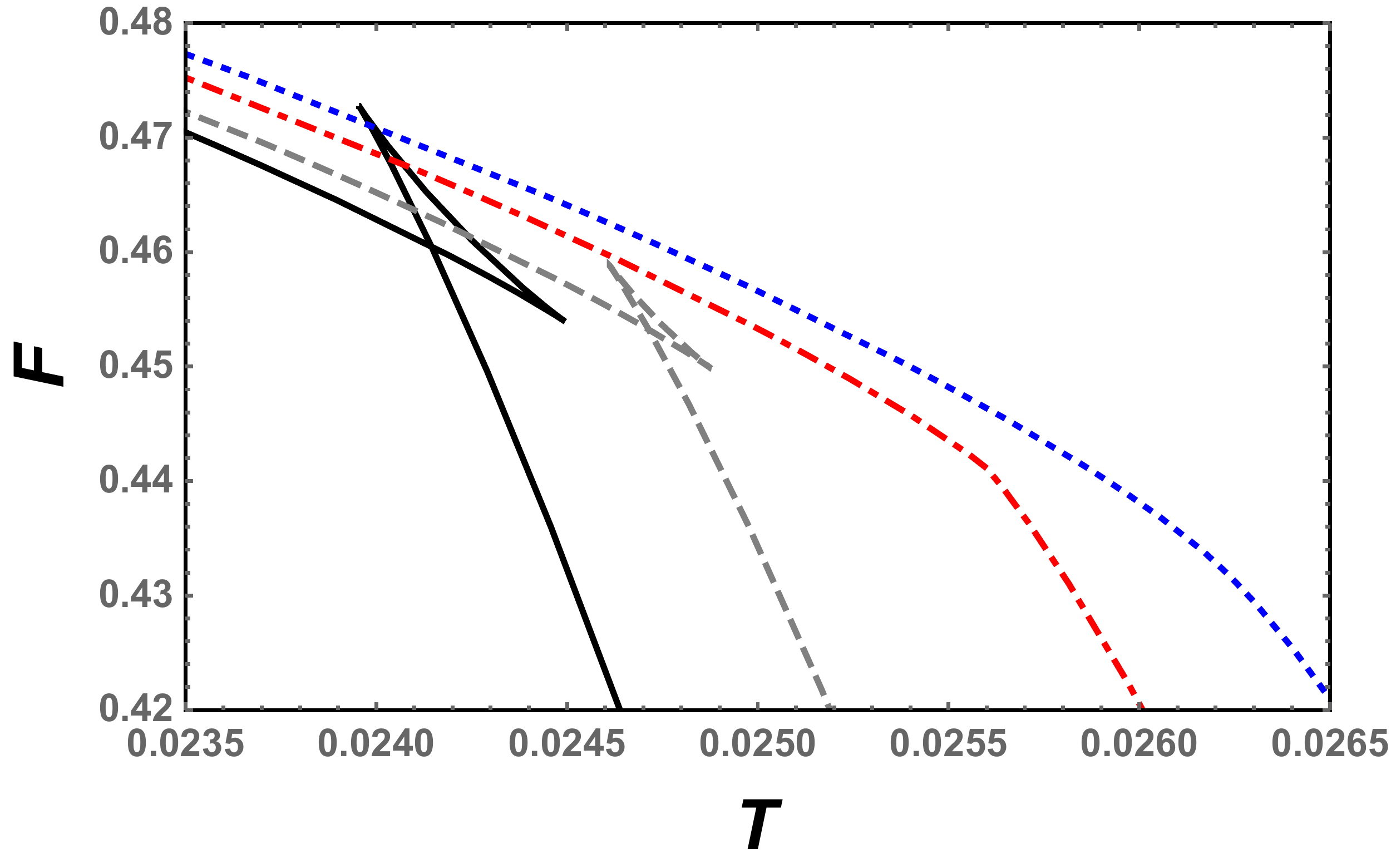}}}
\caption{Isothermal curves and free energy in grand canonical ensemble with $\Phi$=0.5 and $\alpha=0.5$. (a) Isothermal curves for the charged GB-AdS black hole. The temperature $T=0.024, 0.025, 0.02560 (T_c)$, and 0.027 from bottom to top. The black dot at $V=86.57$ denotes the critical point. (b) Free energy $F$ as a function of the temperature $T$ with pressure $P=0.0014, 0.0015, 0.00167 (P_c), 0.0018$ from left to right.}\label{ppGPVa}
\end{figure}

The state equation in the grand canonical ensemble can be expressed as
\begin{eqnarray}
 P=\frac{4 \pi \alpha  T}{3 V}+\frac{\sqrt[3]{\frac{\pi }{6}} (\alpha +3T V)}{3 V^{4/3}}+\frac{\Phi ^2-1}{2\times6^{2/3} \sqrt[3]{\pi } V^{2/3}}.
\end{eqnarray}
The electric potential term is included in the third term, and it clearly modifies the state equation. Here we list the isothermal curves for the charged GB-AdS black hole with $\Phi$=0.5 and $\alpha$=0.5 in Fig. \ref{ppGPV}. For $T$=0.024 and 0.025, we see there are two extremal points on these isothermal curves. While when $T$=0.02560, there exists an inflection point, which corresponds to the critical point of the black hole system. Beyond the value of the temperature, the extremal point disappears. This result is similar to that of the canonical ensemble. So we confirm that there also exists the small-large black hole phase transition in the grand canonical ensemble. This property is different from the charged AdS black hole, where only the Hawking-Page phase transition occurs. Employing the condition (\ref{crcond}), we obtain the corresponding critical point, which reads
\begin{eqnarray}
 P_c&=&\frac{\left(\Phi ^2-1\right)^2 \left(9-3 \Phi
   ^2+\lambda\right)}{24 \pi  \alpha  \left(6-3
   \Phi ^2+\lambda\right)^2},\\
 T_c&=&\frac{\left(8-5 \Phi ^2-\lambda\right)}
    {48 \pi    \alpha }
   \sqrt{\frac{\alpha
   \left(3 \Phi ^2-\lambda-6\right)}
    {{\Phi ^2-1} }},\\
 V_c&=&\frac{4}{3} \pi  \left(\frac{\alpha  \left(3
   \Phi ^2-\lambda-6\right)}{\Phi ^2-1}\right)^{3/2},
\end{eqnarray}
where $\lambda=\sqrt{9 \Phi^4-48 \Phi^2+48}$. From this expression, it is clear that in order to make the critical point to be a physical one, the electric potential must have a range, i.e.,
\begin{eqnarray}
 \Phi\in(0, 1).
\end{eqnarray}
Taking $\alpha$=0, we will have an infinite pressure and temperature and a vanishing volume, which implies that no critical point exists for the charged AdS black hole. On the other hand, if setting the electric potential $\Phi=0$, we will obtain the same result (\ref{ads1})-(\ref{ads2}) as that in the canonical ensemble. The reason is that when the electric potential vanishes, the system is also a neutral GB-AdS black hole system. So the canonical ensemble and the grand canonical ensemble are the same.

Now, the universal quantity constructed in (\ref{univq}) reads
\begin{eqnarray}
 \rho=\frac{1}{48} \left(24-3 \Phi ^2-\lambda\right),\label{uuq}
\end{eqnarray}
which increases with $\Phi$. At $\Phi=0$ and 1, $\rho=\frac{1}{12} \left(6-\sqrt{3}\right)$ and $\frac{3}{8}$, respectively. In this grand canonical ensemble, the corresponding free energy is
\begin{eqnarray}
 F&=&H-TS-Q\Phi\nonumber\\
  &=&-\Phi ^2 r_h+\frac{3 \alpha +8 \pi  P r_h^4+3 \left(\Phi
   ^2+1\right) r_h^2}{6 r_h}\nonumber\\
   &+&\frac{\left(4 \alpha  \ln
   \left(r_h/\sqrt{\alpha}\right)+r_h^2\right) \left(\alpha
   -8 \pi  P r_h^4+\left(\Phi ^2-1\right)
   r_h^2\right)}{4 r_h \left(2 \alpha
   +r_h^2\right)}.\label{fgibbs}
\end{eqnarray}
We display the behavior of the free energy $F$ in Fig. \ref{PPGranFT} with $\Phi=0.5$ and $\alpha=0.5$. Significantly, with the increase of the pressure, the swallow tail behavior tends to disappear. This is a characteristic behavior of the free energy that indicates a first-order small-large black hole phase transition. This phase transition ends at a second-order critical point with the increase of the pressure.

\begin{figure}
\center{\subfigure[]{\label{ppGPTalpha}
\includegraphics[width=7cm]{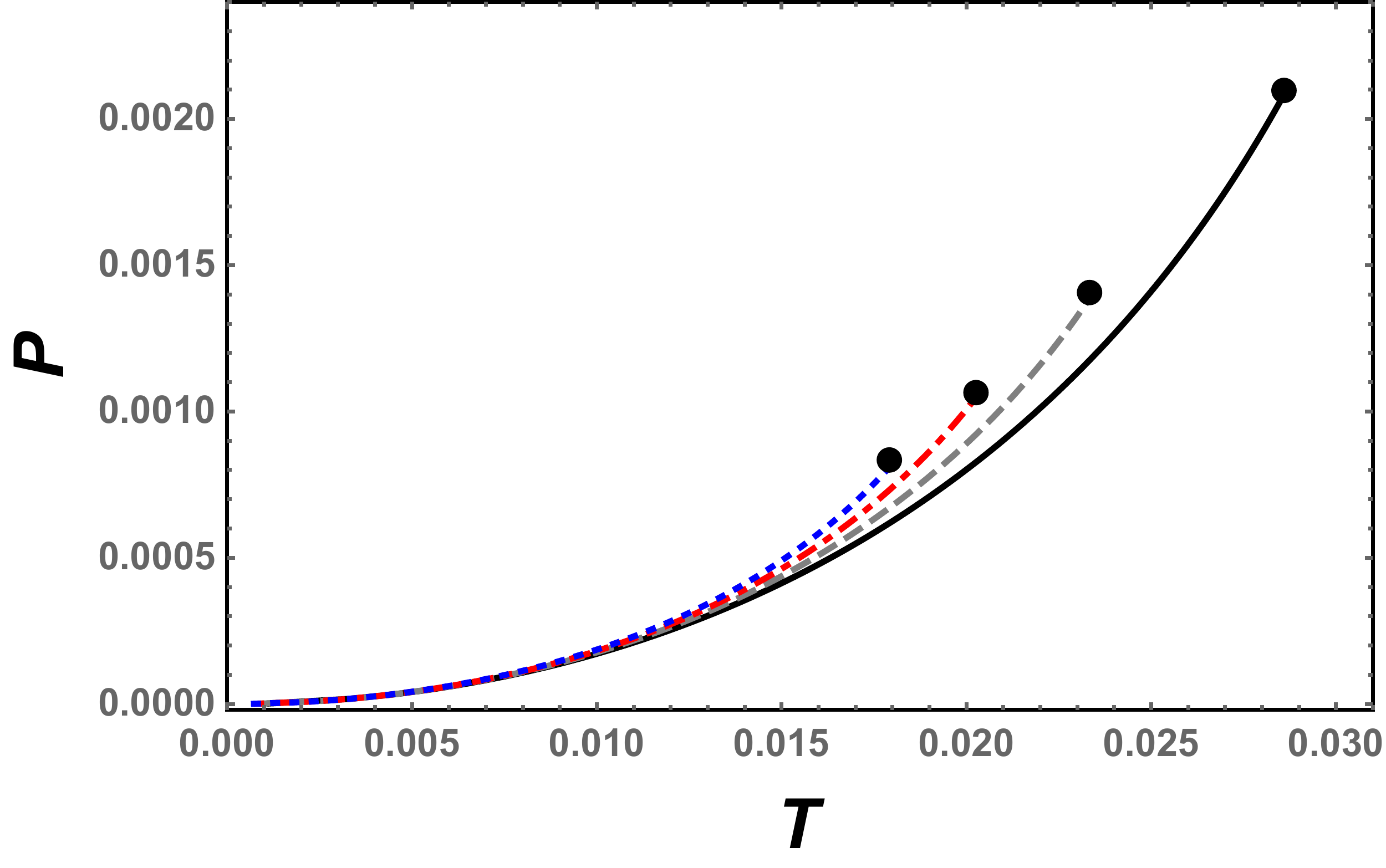}}
\subfigure[]{\label{PPGPTphi}
\includegraphics[width=7cm]{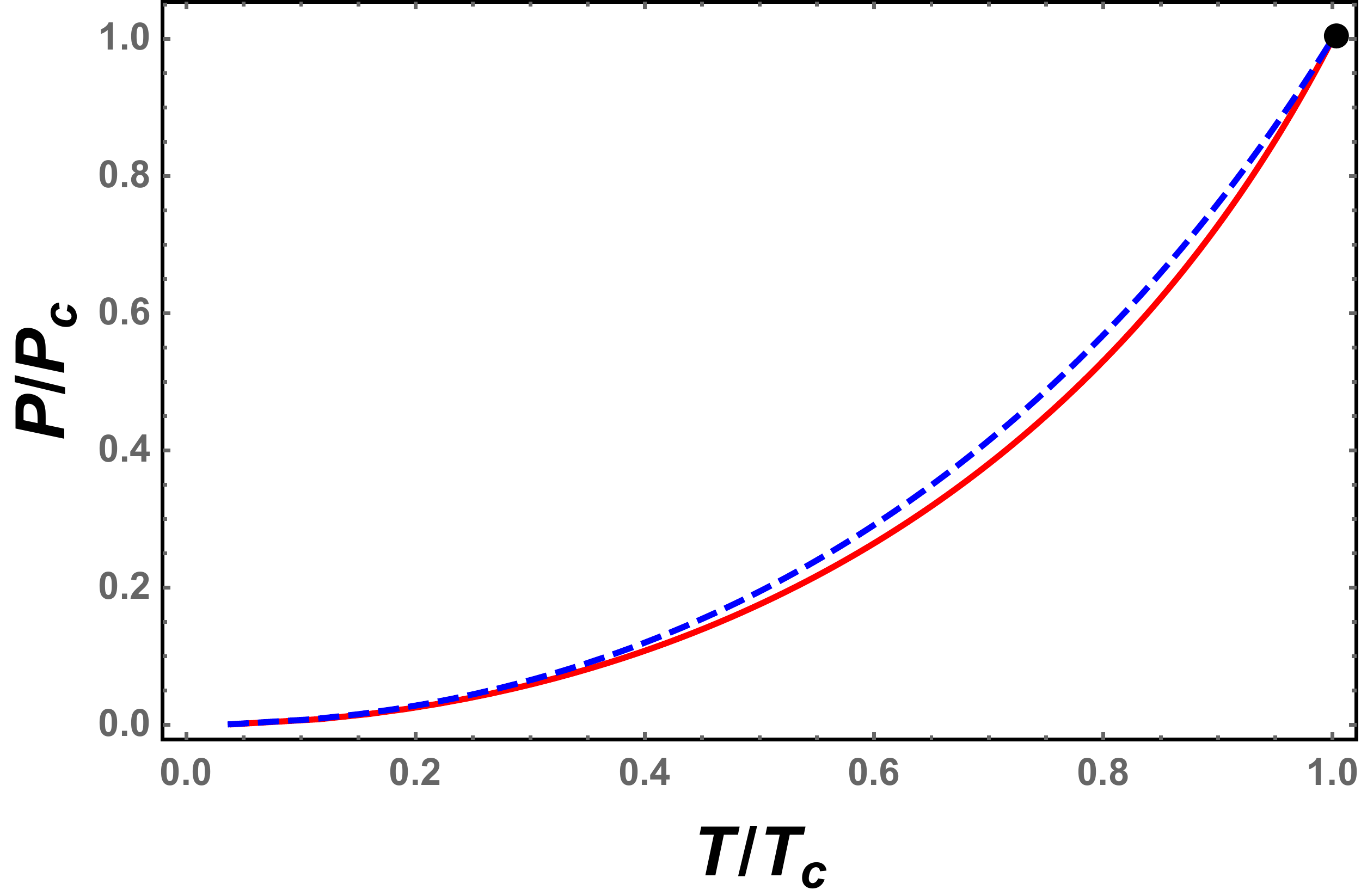}}}
\caption{$P$-$T$ phase diagram for the charged GB-AdS black hole in the grand canonical ensemble. (a) Phase diagram with $\Phi=0.5$ and  $\alpha=0.4, 0.6, 0.8$, and 1.0 from right to left. (b) Phase diagram in the reduced parameter space. The electric potential $\Phi=0$ and 0.99 from bottom to top. In this reduced parameter space, the coexistence is independent of $\alpha$.}\label{ppsGPTphi}
\end{figure}

Making use of the swallow tail behavior of the free energy, we can obtain the coexistence curve. The results are described in Fig. \ref{ppsGPTphi}. Fixing $\Phi=0.5$, we show the coexistence curves in Fig. \ref{ppGPTalpha} for $\alpha=0.4, 0.6, 0.8$, and 1.0 from right to left. Obviously, with the increase of $\alpha$, these curves are shifted to the left. And the critical points are shifted toward to low temperature and pressure. What more interesting is when showing these curves in the reduced $P/P_c$-$T/T_c$ parameter space, we find all these curves coincide with each other. That means for certain $\Phi$, the coexistence curve is independent of $\alpha$. In order to investigate the influence of $\Phi$ on the coexistence curve, we take $\Phi=0$ and 0.99 for example and show these curves in the reduced parameter space, see Fig. \ref{PPGPTphi}. From the figure, we find that the coexistence pressure increases with $\Phi$. However, the influence is very tiny.

Here we would like to discuss why the coexistence curve is independent of $\alpha$ in the reduced parameter space. In Ref. \cite{Weiprd2016}, we introduced the dimensional analysis into the study of the black hole phase transition. This approach has succeeded in obtaining the exact critical point of the phase transition for the $d$-dimensional singly spinning Kerr-AdS black holes \cite{Weiprd2016}, as well as the five-dimensional Kerr-AdS black holes with two equal spin parameters \cite{Weiprd2019}. Although this approach is invalid for the black hole systems with multi-characteristic parameters, it still shows some insight into the double parameters system, see the Kerr-Newman-AdS black hole \cite{Chengprd2016} for an example. This four-dimensional charged GB-AdS black hole is a two-characteristic-parameter system as expected. However, in the grand canonical ensemble, the case becomes a little simple. One key is the electric potential $\Phi$ is dimensionless according to our analysis. Therefore, all these thermodynamic quantities can be reduced by the GB coupling parameter $\alpha$, or alternatively, by their critical values. As a result, all these thermodynamic quantities, as well as the coexistence curve, will only depend on the electric potential $\Phi$. One example is the universal quantity $\rho$ given in (\ref{uuq}), which is obvious independent of $\alpha$. In summary, the reason that the coexistence curve is independent of $\alpha$ in the reduced parameter space is that $\Phi$ is a dimensionless parameter. The analysis here can also be generalized to other higher-dimensional black hole thermodynamics in the grand canonical ensemble.

From above, we see that, in the grand canonical ensemble, there also exists a VdW-like phase transition as in the canonical ensemble. It is also worth to check these critical exponents. For this purpose, one can follow the same process given in the last section. The main treatment is expanding the state equation near the critical point. After the expansion, we get the same expression as (\ref{csel}) while with different coefficients
\begin{eqnarray}
 a_{10}=\frac{4 \left(\lambda +\Phi ^2+2\right)}{11-2
   \Phi ^2},\quad
  a_{11}=\frac{2 \left(-5 \lambda -9 \Phi
   ^2+12\right)}{33-6 \Phi ^2},\quad
   a_{03}=\frac{\lambda +5 \Phi ^2-20}{297-54 \Phi ^2}.
\end{eqnarray}
Since these coefficients do not affect the process, the critical exponents and the scaling laws should be the same as that in the canonical ensemble. So we conclude that the phase transition in the grand canonical ensemble is also the result of the mean field theory.

\section{Black hole microstructures}
\label{bhm}

According to the empirical observation of the well-known thermodynamic geometry, Ruppeiner geometry, a negative or positive scalar curvature always indicates an attractive or repulsive interaction dominating among its microstructures. Moreover, near the critical point, the scalar curvature will go to negative infinity. Thus, the correlation length can also be linked to this scalar. This provides us a powerful approach to insight into the black hole microstructures. Another reason that the geometry attracts so much attention is that it is constructed from the fluctuation theory, which might show us more micro-properties of the back hole systems. Note that these advantages cannot be replaced by other metric geometries. As shown in Ref. \cite{Weiplb}, the five-dimensional neutral GB-AdS black hole has an intriguing microstructure, for which interaction between the microscopic constituents keeps unchanged when the system undergos a small-large black hole phase transition. This motivates us to consider the corresponding properties for the four-dimensional charged GB-AdS black hole. Here, for simplicity, we only focus on the constant charge case.

Following our previous construction, the line element of the Ruppeiner geometry can be expressed with the Gibbs free energy (\ref{gibbs}) as
\begin{equation}
 ds^2=-\frac{1}{T}\left(\frac{\partial^2 \mathcal{F}}{\partial T^2}\right)_{V} dT^2+\frac{1}{T}\left(\frac{\partial^2 \mathcal{F}}{\partial V^2}\right)_{T} dV^2,
\end{equation}
where $\mathcal{F}=H-PV-TS$. We take the temperature $T$ and thermodynamic volume $V$ as the fluctuation coordinates. It should be noted that our original fluctuation coordinates are the mass and volume. Both of them are extensive quantities, which are consistent with the proposal of Ruppeiner geometry \cite{Ruppeiner}. Other recent study on the geometry with different choice of the fluctuation coordinates can be found in Refs. \cite{Xuyang,BhamidipatiGhosh,Guor}.

\begin{figure}
\center{\subfigure[]{\label{ppRNVa}
\includegraphics[width=7cm]{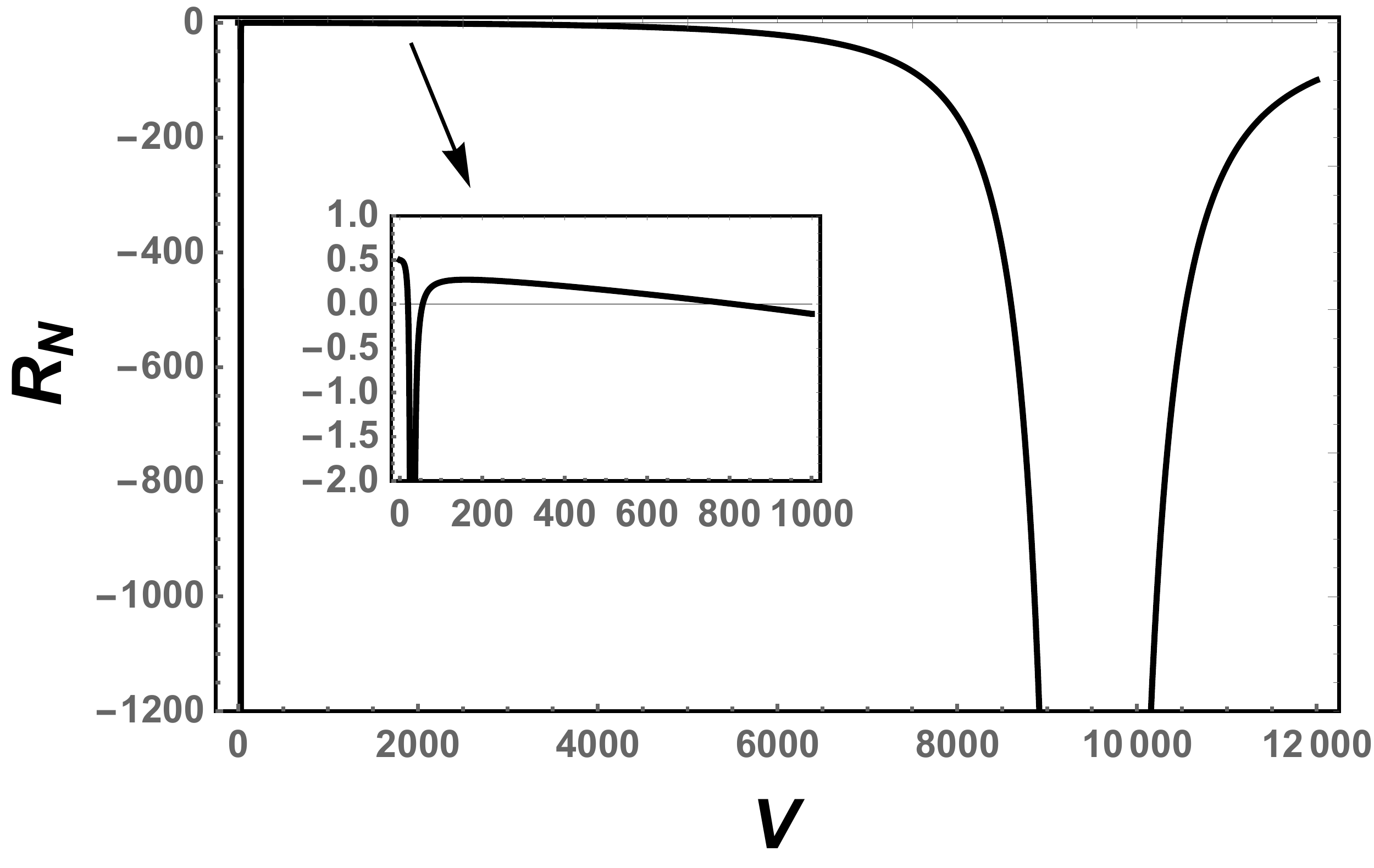}}
\subfigure[]{\label{PPRNVab}
\includegraphics[width=7cm]{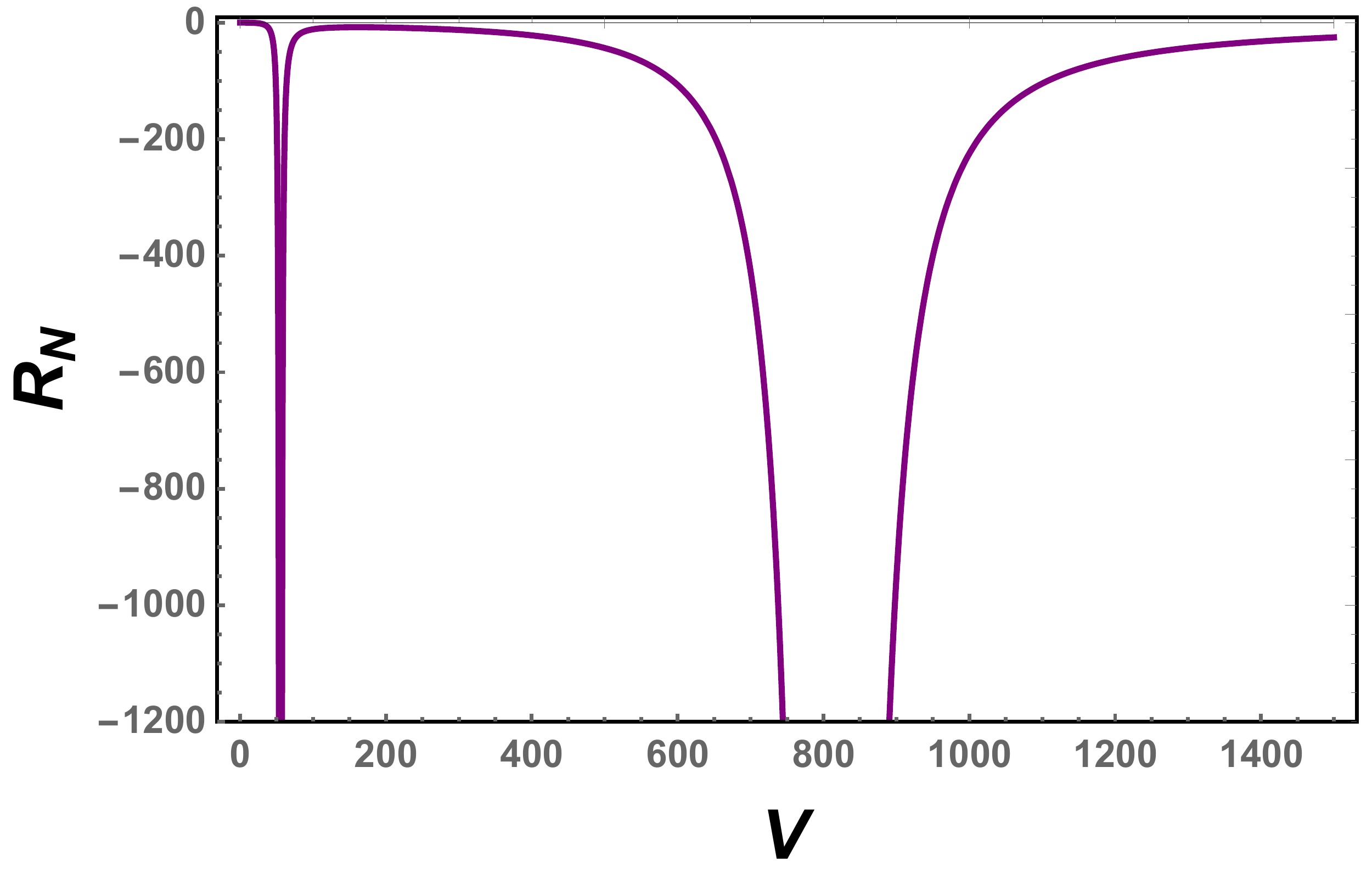}}
\subfigure[]{\label{PPRNVac}
\includegraphics[width=7cm]{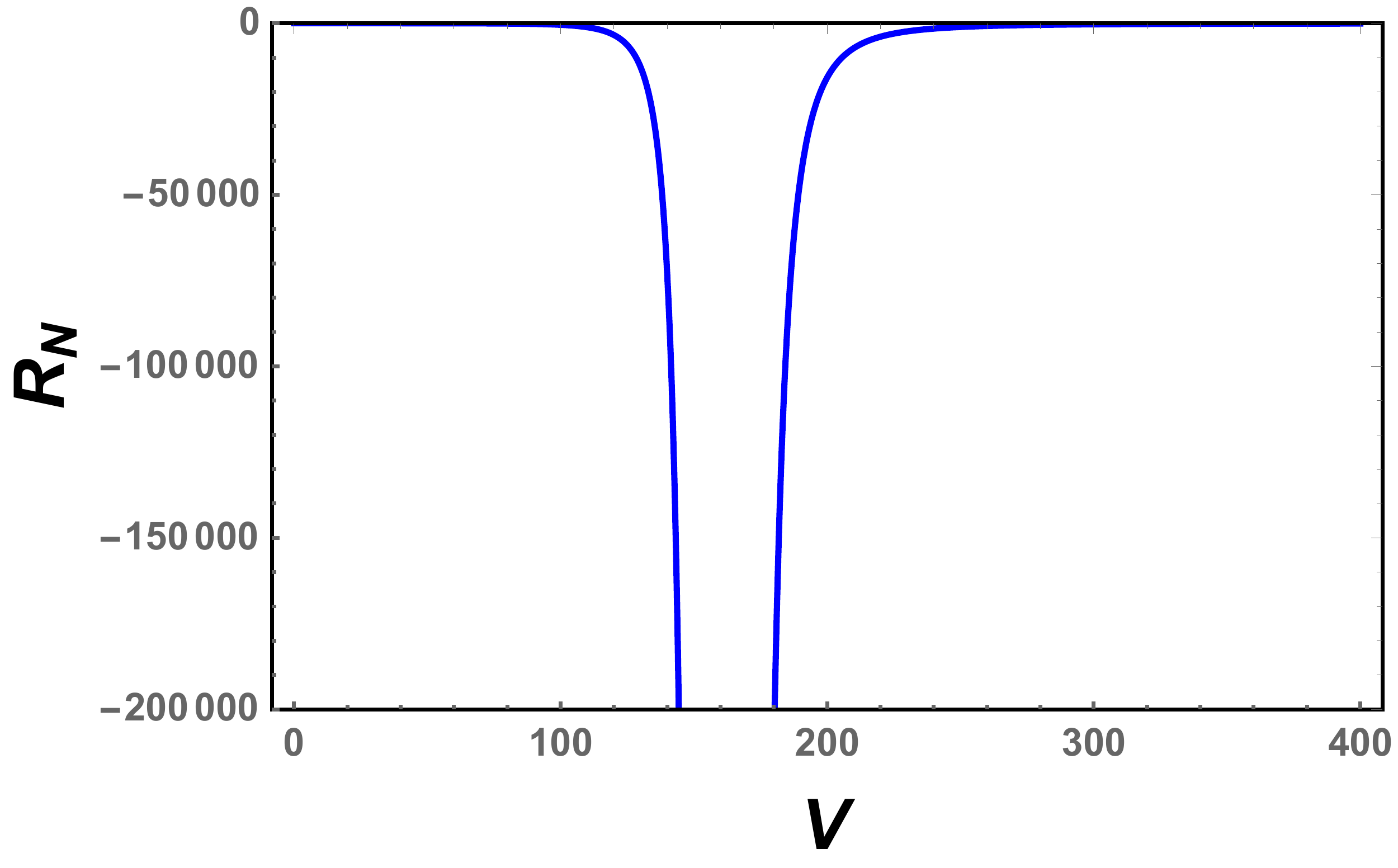}}
\subfigure[]{\label{PPRNVad}
\includegraphics[width=7cm]{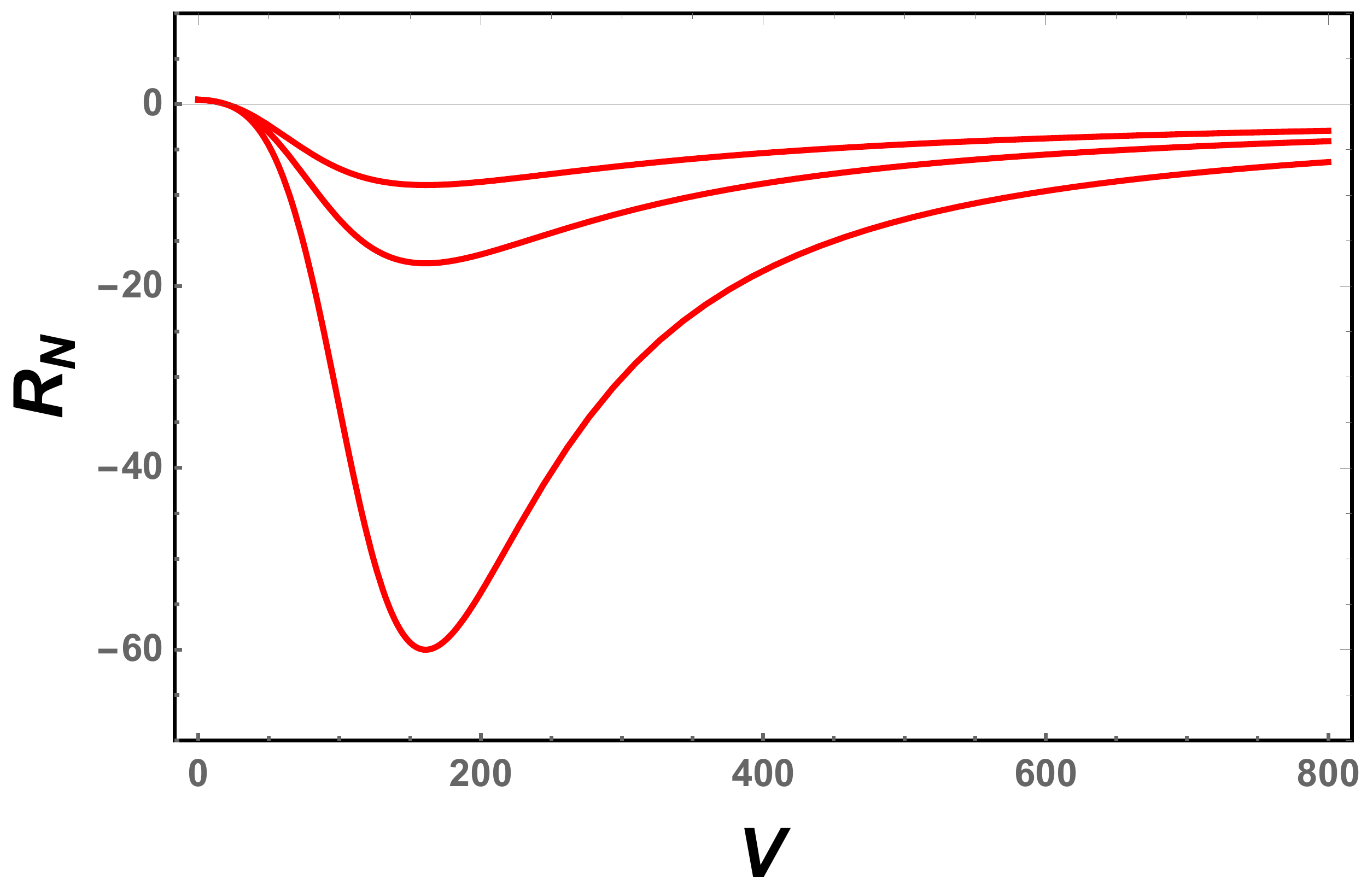}}}
\caption{The normalized scalar curvature $R_N$ as a function of the thermodynamic volume $V$ with $Q$=1 and $\alpha$=0.4 for constant temperature. The corresponding critical temperature $T_c$=0.0294. (a) $T$=0.4$T_c$. (b) $T$=0.8$T_c$. (c) $T$=$T_c$. (d) $T$=1.1$T_c$, 1.2$T_c$ and 1.3$T_c$, from bottom to top.}\label{ppsRNVaphi}
\end{figure}

By using the first law, the line element can be rewritten as
\begin{equation}
 ds^2=-\frac{C_V}{T^2} dT^2+\frac{1}{T}\left(\frac{\partial P}{\partial V}\right)_{T} dV^2.
\end{equation}
Then, the corresponding metric can be easily obtained by the use of the state equation (\ref{ndp}). After a simple calculation, we get the normalized scalar curvature defined in Refs. \cite{Weilm,Weiswliu} for the geometry:
\begin{equation}
 R_N=\frac{1}{2}-\frac{18 T^2 V^{2/3} \left(4 \sqrt[3]{6} \pi ^{2/3} \alpha +V^{2/3}\right)^2}{\left(8 \alpha +8 Q^2+24 \sqrt[3]{6} \pi ^{2/3} \alpha  T
   \sqrt[3]{V}+6 T V-\left(\frac{6}{\pi}\right)^{2/3} V^{2/3}\right)^2}.
\end{equation}
We depict the behavior of $R_{N}$ with $Q$=1 and $\alpha$=0.4 for fixed temperature in Fig. \ref{ppsRNVaphi}. For $T<T_c$, we see that there are two points at which the normalized scalar curvature diverges. While when $T=T_c$, these two divergent points coincide with each other at $V=161$. Further increasing the temperature such that $T>T_c$, no divergent behavior can be found, while only a negative well is presented. The minimum of the well is found to be independent of the black hole temperature.

\begin{figure}
\center{\includegraphics[width=8cm]{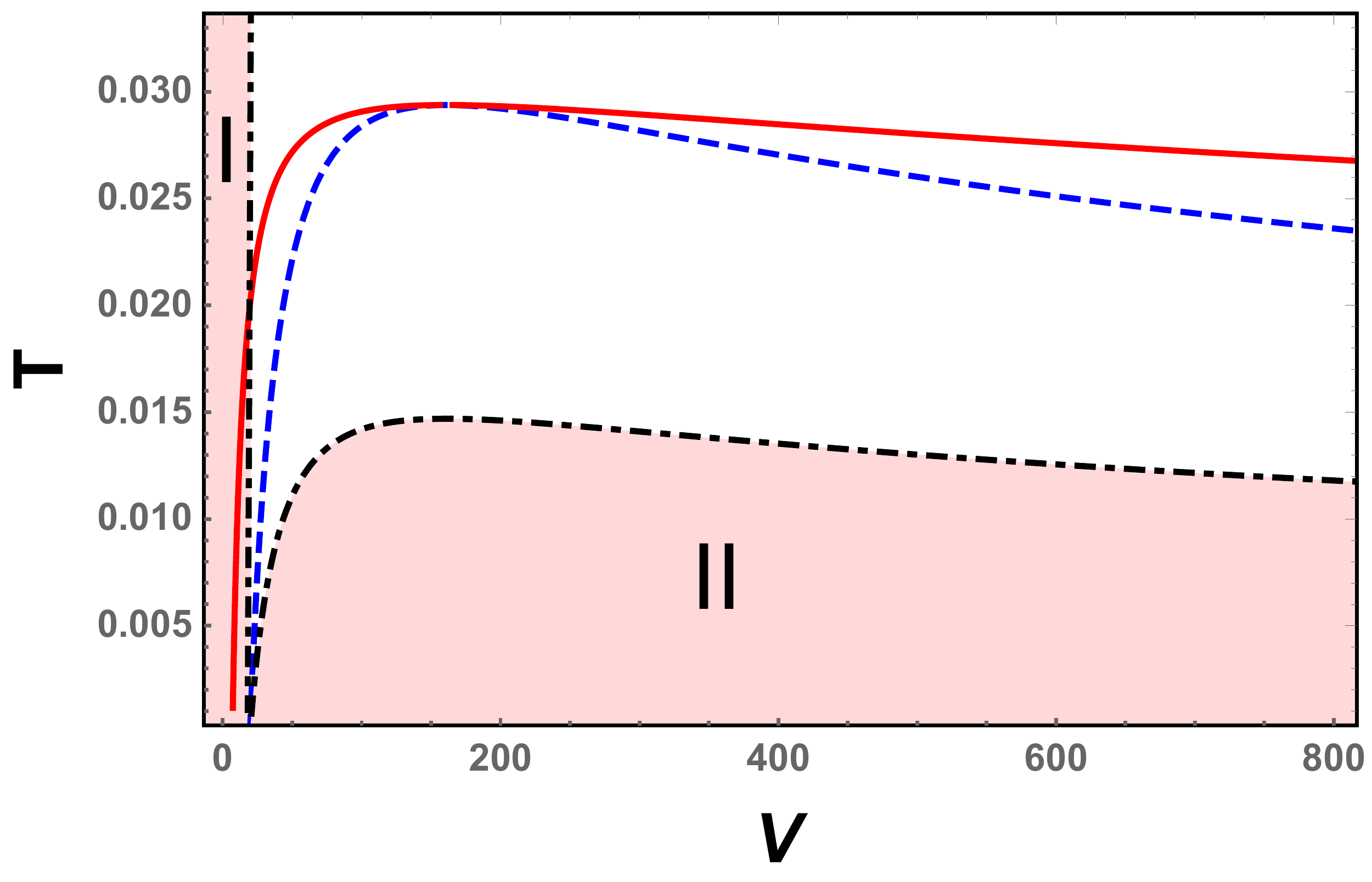}}
\caption{The coexistence curve (red solid curve), spinodal curve (blue dashed curve), and the vanishing $R_N$ (black dot dashed curve). In the shadow regions I and II, $R_N$ is positive, otherwise, it is negative. The parameters are set to $Q=1$ and $\alpha=0.4$.}\label{ppTVc}
\end{figure}

From the empirical observation of the geometry, we know that the most important properties of the scalar curvature are the divergent points and vanishing points. First, solving $1/R_N=0$, we obtain its divergent point
\begin{equation}
 T_{SP}=\frac{\left(\frac{6}{\pi} V \right)^{2/3} -8 \alpha -8 Q^2 }
             {6 \left[4    \alpha ({6\pi^2 V})^{1/3}+V\right]},
\end{equation}
which is just the spinodal curve separating the metastable branch from the unstable branch. For the extremal black hole with vanishing temperature, the thermodynamic volume has a minimum value
\begin{equation}
 V_m=\frac{8}{3} \sqrt{2} \pi  \left(\alpha +Q^2\right)^{3/2}.\label{vm}
\end{equation}
Next, we solve $R_N=0$ and obtain the corresponding temperature $T_0$, which has the following simple relation with $T_{SP}$
\begin{equation}
 T_0=\frac{1}{2}T_{SP}.
\end{equation}
This relation holds for the $d$-dimensional charged AdS black hole \cite{Weilm,Weiswliu}, the five-dimensional neutral GB-AdS black hole \cite{Weiplb}, and the Hayward-AdS black hole \cite{Rizwan}. A natural guess is that this relation holds for all the black hole systems. However, this needs to be further confirmed. On the other hand, at constant volume $V=V_m$, we also have $R_N=0$.

Here we show these curves in Fig. \ref{ppTVc}. This result is very similar to that of the $d$-dimensional charged AdS black hole \cite{Weilm,Weiswliu} and the Hayward-AdS black hole \cite{Rizwan}, while different from the five-dimensional neutral GB-AdS black hole \cite{Weiplb}. In the shadow regions I and II, $R_N$ is positive, otherwise, it is negative. Since in the coexistence region, we do not know whether the state equation holds or not, these results are inapplicable. Excluding the coexistence region, we see region I has positive $R_N$, which implies that the repulsive interaction dominates among the black hole microstructures. While in other region, only the attractive interaction dominates. In summary, significantly different from the five-dimensional neutral GB-AdS black hole, we observe a repulsive microstructure in the four-dimensional charged GB-AdS case for the high temperature small black hole. In addition, from the expression (\ref{vm}), it reveals that such microstructure also exists for the neutral black hole case. Note that when $\alpha=-Q^2$, the four-dimensional system will share the similar behavior as that in five dimensions.

\begin{figure}
\center{\includegraphics[width=8cm]{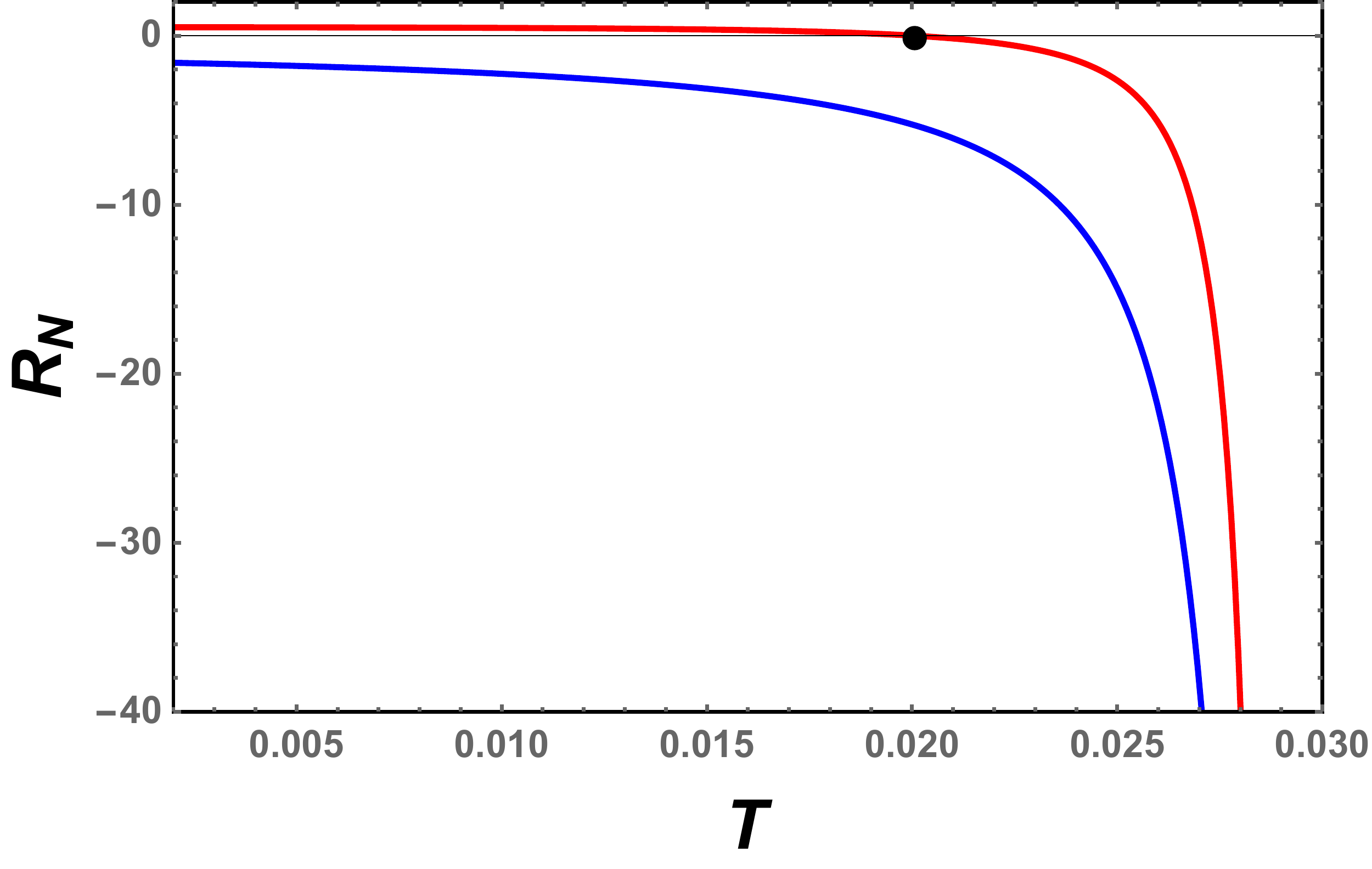}}
\caption{Behaviors of $R_N$ along the small (top red curve) and large (bottom blue curve) black hole coexistence curves. The black dot denotes the vanishing $R_N$ at $T=0.0201$.}\label{ppNTab}
\end{figure}

In the following, we would like to examine the critical behaviors of the normalized scalar curvature $R_N$. As shown in Fig. \ref{ppTVc}, the volumes of the coexistence small and large black holes increase and decrease with the temperature, respectively. At the critical point, the two volumes meet each other and reach the critical volume. Here we plot the behavior of $R_N$ along the coexistence curve in Fig. \ref{ppNTab}. It is clear that both $R_N$ along the coexistence small and large black hole curves decrease with the temperature, and go to negative infinity at the critical point. Moreover, for the coexistence large black hole branch, $R_N$ starts at a negative value. However, for the coexistence small one, $R_N$ is positive at the first, and then turns to negative. These imply that in the low temperature, the coexistence small black hole has a dominant repulsive interaction, while in the high temperature, it has a dominant attractive interaction.

Near the critical point, $R_N$ has a negative divergent behavior like
\begin{equation}
 R_N\sim-|t|^{-a},
\end{equation}
or
\begin{equation}
 \ln |R_N|=-a\ln|t|-b.
\end{equation}
For the four-dimensional charged AdS black hole, $R_N$ has an analytical critical exponent $a=2$. For this black hole system, we numerically fit the coefficients $a$ and $b$ by taking $Q=1$ for different values of $\alpha$. The results are listed in Table \ref{tab1}. From the table, we find that the values of $a$ for the coexistence small and large black holes at the critical points are all round $2$. So for this black hole system, we confirm $R_N$ has a critical exponent 2. Moreover, we find that the parameter $b$ of the coexistence small or large black holes decreases or increases with $\alpha$.

\begin{table}[h]
\begin{center}
\begin{tabular}{cccccc}
  \hline\hline
           & $\alpha$=0.2 & $\alpha$=0.4 & $\alpha$=0.6 & $\alpha$=0.8 & $\alpha$=1.0 \\\hline
 $a$ (CSBH) & 2.0117 & 2.0081 & 2.0058 & 2.0043 & 2.0032\\
 $b$ (CSBH) & 2.2521 & 2.2151 & 2.1924 & 2.1769 & 2.1656 \\\hline
 $a$ (CLBH) & 1.9894 & 1.9917 & 1.9932 & 1.9943 & 1.9950  \\
 $b$ (CLBH) & 1.9076 & 1.9345 & 1.9513 & 1.9630 & 1.9630  \\ \hline\hline
\end{tabular}
\caption{Fitting values of $a$ and $b$ for the coexistence small black hole (CSBH) and coexistence large black hole (CLBH) with $Q=1$ and $\alpha=0.4$.}\label{tab1}
\end{center}
\end{table}

Furthermore, with these numerical results given in Table \ref{tab1}, we obtain another universal quantity $R_Nt^2$ with the values given in Table \ref{tab2}. For the charge $Q=1$, the absolute value of $R_Nt^2$ slightly increases with $\alpha$. For the four-dimensional charged AdS and neutral GB-AdS black holes, they share an analytical value $-\frac{1}{8}$, which also coincides with that of small $\alpha$.

\begin{table}[h]
\begin{center}
\begin{tabular}{cccccc}
  \hline\hline
           & $\alpha$=0.2 & $\alpha$=0.4 & $\alpha$=0.6 & $\alpha$=0.8 & $\alpha$=1.0 \\\hline
 $R_Nt^2$ & -0.1250 & -0.1256 & -0.1260 & -0.1262 & -0.1264 \\ \hline\hline
\end{tabular}
\caption{Universal values of $R_Nt^2$ near the critical point with $Q=1$.}\label{tab2}
\end{center}
\end{table}

\section{Discussions and conclusions}
\label{Conclusion}

In this paper, we studied the extended thermodynamics of the four-dimensional charged GB-AdS black hole both in the canonical ensemble and grand canonical ensemble. Its microstructure was also investigated.

At first we examined the first law of thermodynamics for the four-dimensional charged GB-AdS black hole in the extended phase space. The cosmological constant was treated as pressure and its conjugate quantity as the thermodynamic volume, which equals the value of the geometrical volume inside the horizon. However, they are two different concepts. Besides, the GB coupling $\alpha$ was also treated as a thermodynamic variable. Then the first law and the corresponding Smarr formula were found to hold for the four-dimensional charged GB-AdS black hole.

In the canonical ensemble where the charge is fixed, we found a small-large black hole phase transition with a critical point. The coexistence curve in the $P$-$T$ diagram was shown, which is like that of the VdW fluid. Moreover, near the critical point, we calculated the critical exponents by using the state equation, which is the same with the VdW fluid. In order to determine the phase transition point, one can use the Maxwell equal area law or the Gibbs free energy. Here we pointed out that the results obtained by these two methods only coincide if the first law holds. Since in the GB gravity, the entropy depends on $\alpha$, one should be very careful in choosing the specific form of the entropy, especially the integral constant. These will affect the Gibbs free energy, as well as the phase transition point. On the other hand, if we consider the case with fixed charge and GB coupling, then constructing the equal area law on each isothermal curve will produce the right phase transition point.

When excluding the pure thermal radiation phase, we also observed a VdW-like phase transition in the grand canonical ensemble, where the electric potential is fixed. The critical point and the coexistence curve were obtained. In the reduced parameter space, we found that all the thermodynamic quantities and the coexistence curve are independent of the GB coupling parameter. This can be understood with the dimensional analysis we proposed before. Due to the fact that the electric potential is dimensionless, then in the reduced parameter space, the thermodynamics will only show an electric potential dependent behavior for the four-dimensional charged GB-AdS black hole. Performing the same calculation in the canonical ensemble, we obtained the same critical exponents and scaling law. One thing worth to point out is that for the absence of the charge or electric potential, there still exists a small-large black hole phase transition of the VdW type.

After considering the thermodynamics and phase transition, we turned to the study of the black hole microstructures by using the Ruppeiner geometry. We calculated the normalized scalar curvature, which is believed to contain the information of the dominant interaction in its sign. Near the critical point, it is also linked to the correlation length. For fixing charge and GB coupling parameter, we observed two divergent points of the scalar curvature for low temperature. These two points coincide at the critical point. While beyond the critical point, only a negative well is presented with its minimum locating at constant volume.

By solving the normalized scalar curvature, we obtained the temperatures $T_{SP}$ and $T_0$ of the divergent point and the vanishing point. They satisfy the simple relation $T_0=\frac{1}{2}T_{SP}$ as other black hole systems do. As we know, in the coexistence region, the state equation may not hold any more, and thus we excluded that region. We found that the survival region I still has positive scalar curvature, which indicates there is a dominant repulsive interaction. This result is rather different from the five-dimensional neutral GB-AdS black hole, where only the dominant attractive interaction appears. Further, when taking a zero charge, we got a non-vanishing $V_m$ (\ref{vm}), which indicates that the structure still keeps unchanged, see Fig. \ref{ppTVc} for the neutral case. Therefore, the dominant repulsive interaction structure is universal for the four-dimensional GB-AdS black hole. This result provides us a preliminary microstructure knowledge of the GB gravity in four and higher dimensions.

Moreover, we numerically calculated the coexistence curve near the critical point, and fitted the coefficients for the scalar curvature. The result confirms that the normalized scalar curvature has a characteristic universal critical exponent 2. Adopting the fitting coefficients, we got $R_Nt^2$. With the increase of $\alpha$, its absolute value slightly increases.

In conclusion, the four-dimensional GB gravity has significant difference from the higher-dimensional one. Our results on the black hole phase transition and the microstructures will provide further insight into the four-dimensional GB gravity.

\section*{Acknowledgements}

This work was supported by the National Natural Science Foundation of China (Grants No. 11675064 and No. 11875151).

\end{document}